\documentclass[nofootinbib,showpacs,superscriptaddress,
preprintnumbers
,prd,aps]{revtex4}
\usepackage{graphicx,amsmath,amssymb,dcolumn,bm}
\usepackage[utf8]{inputenc}
\usepackage{tensor}
\usepackage{kbordermatrix}
\usepackage{rotating}
\usepackage{xcolor}
\usepackage{hyperref}
\hypersetup{
    colorlinks,
    linkcolor={red!50!black},
    citecolor={blue!50!black},
    urlcolor={blue!80!black}
}

\newcommand{\mc}[1]{\mathcal{#1}}
\newcommand{\bra}[1]{\langle{#1}|}
\newcommand{\ket}[1]{|{#1}\rangle}
\newcommand{\braket}[2]{\langle{#1}|{#2}\rangle}

\begin{document}
\title{Transversity generalized parton distributions for the deuteron}
\author{W. Cosyn}
\affiliation{Department of Physics and Astronomy, Ghent University, 
             Proeftuinstraat 86, B9000 Ghent, Belgium}
\author{B. Pire}
\affiliation{Centre de Physique Th\'{e}orique, \'{E}cole Polytechnique, CNRS, 
91128 Palaiseau, France}
\date{\today}
\begin{abstract}
Transversity generalized parton distributions (GPDs) appear as scalar functions 
in the decomposition of off-forward quark-quark and gluon-gluon correlators with a parton 
helicity flip.  For a spin 1 hadron, we find 9 transversity GPDs for both quarks 
and gluons at leading twist 2.  We study these twist-2 chiral odd quark 
transversity GPDs for the deuteron in a light cone convolution model, based on 
the impulse approximation, and using the lowest Fock-space state for the deuteron.  
\end{abstract}
\maketitle

\section{Introduction}
\label{sec:intro}
The factorization of hard exclusive amplitudes in the generalized Bjorken regime~\cite{Diehl:2003ny, Belitsky:2005qn} as the convolution of  generalized parton distributions (GPDs) with  perturbatively calculable coefficient functions allows to get access to the $3-$dimensional structure of nucleons or nuclei through the extraction of the various quark and gluon GPDs. The connection between GPDs and parton-hadron helicity amplitudes allows an easy counting of twist-2 GPDs :  there  are $2(2J+1)^2$ GPDs for each quark flavor (or for the gluon) in a nucleus of spin $J$. Half of these GPDs correspond to parton helicity non-flip, the other half - which are dubbed {\em transversity GPDs} - correspond to parton helicity flip. In the quark case, the helicity non-flip GPDs are chiral even, while the helicity flip GPDs are chiral-odd.  The helicity flip and non-flip sectors evolve independently in the renormalization scale. Moreover, the quark and gluon sectors do not mix in the evolution of transversity GPDs.

Nuclear GPDs \cite{Fucini:2018gso, Dupre:2015jha, Rinaldi:2012pj, Taneja:2011sy, Scopetta:2009sn, Liuti:2005gi, Liuti:2005qj, Freund:2003ix, Guzey:2003jh, Scopetta:2004kj} obey the same rules as nucleon GPDs and are accessible through coherent exclusive processes which may be isolated from incoherent processes where the target nucleus breaks during the hard interaction.  
As the simplest composite nucleus, the deuteron is a fascinating object to scrutinize in order to understand the QCD confinement mechanism~\cite{Boeglin:2015cha}. The study of hard reactions which allow to access its quark and gluon structure is at the heart of the on-going physics program at Jefferson Lab (JLab) as well as the future electron-ion collider (EIC) program. The study of the deuteron GPDs should allow to understand more deeply the relation between the deuteron and nucleon structures.  The spin 1 nature of the deuteron makes it a particularly rich object from the point of view of building the spin from the constituent spins and orbital angular momenta. 

Contrarily to the nucleon GPDs which have been the subject of many works -- both theoretically and experimentally -- the study of deuteron GPDs is still in its infancy; its founding blocks are the definition of helicity non-flip quark and gluon GPDs ~\cite{Berger:2001zb} and the calculation of deeply virtual Compton scattering (DVCS) and  deep exclusive meson production  (DEMP) amplitudes \cite{Kirchner:2003wt, Cano:2003ju} in the coherent reactions on a deuteron. First results on coherent hard exclusive reactions have been obtained at JLab \cite{Mazouz:2017skh}. In the present paper, we study the transversity sector of deuteron twist-2 GPDs which was left aside up to now.

The  paper is organized as follows.  The transversity GPDs of spin 1 hadrons are the objects of study in Sec.~\ref{sec:gpd_t}: we start with introducing the necessary kinematic variables in Subsec.~\ref{subsec:kin}, list the general correlators and their symmetry properties in Subsec.~\ref{subsec:gpd_rel}, and subsequently introduce the transversity GPDs for spin 1 and comment on their properties for quarks (Subsec.~\ref{subsec:gpd_t_q}) and gluons (Subsec.~\ref{subsec:gpd_t_g}).  In the following Section~\ref{sec:d_conv_form}, we outline the convolution formalism for the deuteron, with kinematic variables defined in Subsec.~\ref{subsec:conv_kin}, the deuteron light-front wave function and chiral odd nucleon GPDs discussed in Subsecs.~\ref{subsec:d_lf_wf} and \ref{subsec:nucl_gpd}, and finally the convolution model is presented in Subsec.~\ref{subsec:conv_derivation}.  Results obtained in the convolution formalism for transversity helicity amplitudes and GPDs in the quark sector are discussed in Sec.~\ref{sec:results} and sum rules from the first moments of the quark transversity GPDs are covered in Sec.~\ref{sec:sumrules}.  Conclusions are stated in Sec.~\ref{sec:concl}.  The notation, sign and normalization conventions used throughout this article are summarized in Appendix~\ref{app:conv}, while App.~\ref{app:discrete} contains a summary of the properties of parity and time reversal symmetries on the light front.  The relations between transversity helicity amplitudes and GPDs for spin 1 hadrons are listed in App.~\ref{app:helamps}, and a minimal convolution model used to obtain some analytical results is outlined in App.~\ref{app:min_conv}.

We shall not  deal with the phenomenology of these GPDs in this article, and leave this topic for further work.  Similarly, the polynomiality properties of spin 1 GPDs and the connection between general moments and the generalized form factors will be discussed elsewhere.  At present, no parameterization for the nucleon gluon transversity GPDs is available~\cite{Pire:2014fwa}.  Consequently, in this article we do not consider calculations in the convolution model for the deuteron gluon transversity GPDs.

\section{Transversity GPDs for spin 1 hadrons}
\label{sec:gpd_t}
\begin{figure}[htb]
\begin{center}
 \includegraphics[width=0.4\textwidth]{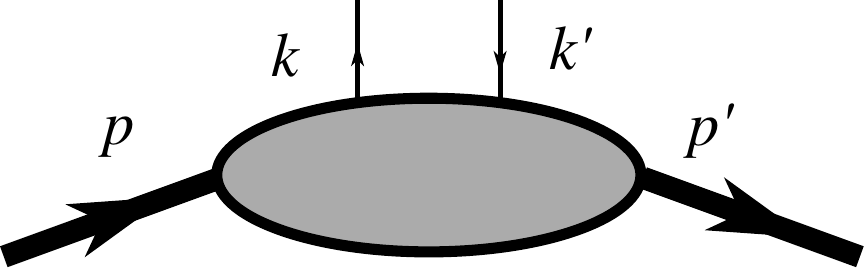}
 \end{center}
 \caption{Diagrammatic representation of an off-forward parton correlator.
 \label{fig:corr}}
\end{figure}

The central objects that define GPDs are Fourier transforms of gauge-invariant off-forward parton correlators, where the initial (final) hadron in the correlator matrix element has four-momentum $p$ ($p'$), light-front helicity $\lambda$ ($\lambda'$) and mass $M$.  For quarks these correlators take the form
\begin{equation}
\bra{p'\,\lambda'}\bar{\psi}(-\kappa 
n)\Gamma \psi(\kappa n)\ket{p\,\lambda}\,,
\end{equation}
with $\Gamma$ a general Dirac structure, and the two quark fields are separated along a light-like four-vector $n^\mu$ ($n^2=0$).  In this work, we use the lightcone gauge $(nA)=0$, so no explicit Wilson lines appear in the correlators.  Similar correlators can also be introduced for gluons (see Subsec.~\ref{subsec:gpd_rel}).  These objects encode long distance, strongly coupled QCD dynamics and can be diagramatically represented by the blob in Fig.~\ref{fig:corr}.

\subsection{Kinematical variables}
\label{subsec:kin}

We introduce the standard kinematic variables for these matrix elements, being the  average hadron momentum $P$, momentum transfer $\Delta$, skewness $\xi$ (which determines the longitudinal momentum transfer) and $t$:
\begin{align}\label{eq:kin_d}
 &P=\frac{p+p'}{2}\,,
 \nonumber\\
  &\Delta=p'-p\,,\qquad t=\Delta^2\,,\nonumber\\ 
  &\xi=-\frac{(\Delta 
n)}{2(Pn)}\,.
\end{align}
Depending on the skewness $\xi$, the momentum transfer squared $t$ (which is negative) has a maximum value 
\begin{equation}
t_0=-\frac{4M^2\xi^2}{1-\xi^2}\,,
\end{equation}
and we can write
\begin{equation}
t_0-t=-2M^2\frac{1+\xi^2}{1-\xi^2}+2(pp')\,.
\end{equation}

The four-vector $2\xi P+\Delta$ is orthogonal to $n$ $\left[((2\xi P+\Delta)n)=0\right]$ and has norm
\begin{equation}
(2\xi P+\Delta)^2=-(1-\xi^2)(t_0-t) \,.
\end{equation}
 
The following combination of kinematic variables occurs a lot in formulas in this work, 
so an extra dimensionless variable is defined:
\begin{equation}
 D\equiv \frac{\sqrt{(t_0-t)(1-\xi^2)}}{2M}\,.
\end{equation}

As we study parton correlators for spin 1 particles, we consider a basis of three polarization four-vectors, both 
for the initial (unprimed four-vectors) and final (primed four-vectors) spin 1 hadron state~\cite{Berger:2001zb}, normalized to $(\epsilon^{*(i)}\epsilon^{(i)})=-1$ and orthogonal to the particle four-momentum [$(\epsilon(i)p)=(\epsilon
'(i)p')=0$]~\footnote{Our sign convention for the Levi-Civita tensor and other quantities is summarized in App.~\ref{app:conv}}:
\begin{align}\label{eq:pol_vectors}
 &\epsilon^{(0)\mu}=\frac{1}{M}\left(p^\mu-\frac{M^2}{1+\xi}\frac{n^\mu}{(Pn)} 
\right)\,,\nonumber\\
&\epsilon'^{(0)\mu}=\frac{1}{M}\left(p'^\mu-\frac{M^2}{1-\xi}\frac{n^\mu}{ (Pn) 
} \right)\,, \nonumber\\
&\epsilon^{(1)\mu}=-\frac{1}{\sqrt{(1-\xi^2)(t_0-t)}}
\left((1+\xi)p'^\mu-(1-\xi)p^\mu -\frac{\xi(t_0-t)-t_0}{2\xi}\frac{n^\mu}{(Pn)} 
\right)\,,\nonumber\\
&\epsilon'^{(1)\mu}=-\frac{1}{\sqrt{(1-\xi^2)(t_0-t)}}
\left((1+\xi)p'^\mu-(1-\xi)p^\mu +\frac{\xi(t_0-t)+t_0}{2\xi}\frac{n^\mu}{(Pn)} 
\right)\,,\nonumber\\
&\epsilon^{(2)\mu}=\epsilon'^{(2)\mu}=\frac{1}{\sqrt{(1-\xi^2)(t_0-t)}}
\frac{\tensor{\epsilon}{^\mu_{\nu\alpha\beta}}\;p'^{\nu} p^\alpha 
n^\beta}{(Pn)}\,.
\end{align}
We use
\begin{align}\label{eq:pol_lfhel}
&\epsilon(0)=\epsilon^{(0)}\,,\nonumber\\
&\epsilon(\pm)=\mp e^{\pm i\phi}(\epsilon^{(1)}\pm i \epsilon^{(2)})/{\sqrt{2}}\,,
\end{align}
 as 
definite light-cone helicity polarization four-vectors for the initial hadron, and similar expressions for the primed polarization four-vectors and the final hadron.  In Eq.~(\ref{eq:pol_lfhel}), $\phi$ is the 
azimuthal angle of the four-vector $\Delta+2\xi P$.


\subsection{Correlators and symmetry properties}
\label{subsec:gpd_rel}
The following quark-quark correlators determine the leading twist-2 quark GPDs~\cite{Belitsky:2005qn,Diehl:2003ny,Diehl:2001pm}
\begin{align}
 V^q_{\lambda'\lambda}&=\int 
\frac{d\kappa}{2\pi}e^{2ix\kappa(Pn)}\bra{p'\,\lambda'}\bar{\psi}(-\kappa 
n)(\gamma n)\psi(\kappa n)\ket{p\,\lambda}\,,\nonumber\\
 A^q_{\lambda'\lambda}&=\int 
\frac{d\kappa}{2\pi}e^{2ix\kappa(Pn)}\bra{p'\,\lambda'}\bar{\psi}(-\kappa 
n)\gamma_5(\gamma n)\psi(\kappa n)\ket{p\,\lambda}\,,\nonumber\\
 T^{q\,i}_{\lambda'\lambda}&=\int 
\frac{d\kappa}{2\pi}e^{2ix\kappa(Pn)}\bra{p'\,\lambda'}\bar{\psi}(-\kappa 
n)(in_\mu\sigma^{\mu i})\psi(\kappa n)\ket{p\,\lambda}\,,
\label{eq:corr_def}
\end{align}
where $i$ is a transverse 
index and transverse is relative to the light-like four-vectors $n$ and $\bar{n}$ ($n^2=\bar{n}^2=0$, $n\bar{n}$=1).  The decomposition for the first two (vector $V^q_{\lambda'\lambda}$, axial vector $A^q_{\lambda'\lambda}$) was considered for spin 1 hadrons in Ref.~\cite{Berger:2001zb} and determines the 9 chiral even quark GPDs  for spin 1 (5 for $V^q_{\lambda'\lambda}$, 4 for $A^q_{\lambda'\lambda}$).  The decomposition of the tensor correlator $T^{q\,i}_{\lambda'\lambda}$ is given below (Subsec.~\ref{subsec:gpd_t_q}) and determines 9 spin 1 chiral odd quark GPDs.
 
Similarly, the following gluon-gluon correlators determine the leading twist-2 gluon GPDs~\cite{Belitsky:2005qn,Diehl:2003ny,Diehl:2001pm}:
\begin{align}\label{eq:corr_def_g}
 V^g_{\lambda'\lambda}&=\frac{2}{(Pn)}\int 
\frac{d\kappa}{2\pi}e^{2ix\kappa(Pn)}\bra{p'\,\lambda'}\text{Tr}\left[ n_\alpha 
G^{\alpha \mu}(-\kappa n) 
\tensor{G}{_\mu^\beta}(\kappa n)n_\beta\right]\ket{p\,\lambda} \nonumber\\
&=\frac{1}{(Pn)}\int 
\frac{d\kappa}{2\pi}e^{2ix\kappa(Pn)}\bra{p'\,\lambda'}\text{Tr}\left[n_\alpha 
\left( G^{\alpha R}(-\kappa n) 
G^{\beta L}(\kappa n)+G^{\alpha L}(-\kappa n) 
G^{\beta R}(\kappa n)\right) n_\beta\right]\ket{p\,\lambda}\,,
\nonumber\\
 A^g_{\lambda'\lambda}&=-\frac{2i}{(Pn)}\int 
\frac{d\kappa}{2\pi}e^{2ix\kappa(Pn)}\bra{p'\,\lambda'}\text{Tr}\left[n_\alpha 
G^{\alpha \mu}(-\kappa n) 
\tensor{\widetilde{G}}{_\mu^\beta}(\kappa n)n_\beta\right]\ket{p\,\lambda}
\nonumber\\
&=\frac{1}{(Pn)}\int 
\frac{d\kappa}{2\pi}e^{2ix\kappa(Pn)}\bra{p'\,\lambda'}\text{Tr}\left[n_\alpha 
\left( G^{\alpha R}(-\kappa n) 
G^{\beta L}(\kappa n)-G^{\alpha L}(-\kappa n) 
G^{\beta R}(\kappa n)\right) n_\beta\right]\ket{p\,\lambda}\,,
\nonumber\\
 T^{g\,ij}_{\lambda'\lambda}&=-\frac{2}{(Pn)}\int 
\frac{d\kappa}{2\pi}e^{2ix\kappa(Pn)}\bra{p'\,\lambda'}\text{Tr}\hat{\bm{S}
}\left[
 n_\alpha 
G^{\alpha i}(-\kappa n) n_\beta G^{\beta j}(\kappa n) 
\right]\ket{p\,\lambda}\, 
,
\end{align}
where $i,j$ are  transverse 
indices, the operator $\hat{\bm S}$ implies symmetrisation and removal of 
trace, and transverse four-vector components $a^{R/L}$ are defined as
\begin{align}
 &a^R=a^x+ia^y\,,\nonumber\\
  &a^L=a^x-ia^y\,.
\end{align}
Again, the decomposition of $V^q_{\lambda'\lambda},A^q_{\lambda'\lambda}$ for spin 1 hadrons has been discussed earlier~\cite{Berger:2001zb} and the composition of the tensor correlator $T^{g\,ij}_{\lambda'\lambda}$ is given below in Subsec.~\ref{subsec:gpd_t_g}.

As $T^{g\,RL}_{\lambda'\lambda}= T^{g\,LR}_{\lambda'\lambda}=0$, there remain two independent matrix 
elements for the tensor gluon-gluon correlator:
\begin{align}
 T^{g\,RR}_{\lambda'\lambda}&=-\frac{2}{(Pn)}\int 
\frac{d\kappa}{2\pi}e^{2ix\kappa(Pn)}\bra{p'\,\lambda'}\text{Tr}\left[
 n_\alpha 
G^{\alpha R}(-\kappa n) n_\beta G^{\beta R} (\kappa n)\right]\ket{p\,\lambda}\, 
, 
\nonumber\\
 T^{g\,LL}_{\lambda'\lambda}&=-\frac{2}{(Pn)}\int 
\frac{d\kappa}{2\pi}e^{2ix\kappa(Pn)}\bra{p'\,\lambda'}\text{Tr}\left[
 n_\alpha 
G^{\alpha L}(-\kappa n) n_\beta G^{\beta L}(\kappa n) \right]\ket{p\,\lambda}\, 
. 
\end{align}

Hermiticity and discrete light-front symmetries~\footnote{The properties of light-front parity and time reversal are summarized in App.~\ref{app:discrete}} impose the following 
constraints on the correlators~\footnote{If the correlators in the following equations do not 
have a $q$ or $g$ superscript, the same relation 
is valid for both the quark-quark and gluon-gluon correlator.  Transverse superscripts separated by a slash denote multiple possible values to be considered in sequence between the left- and right-hand side.}:
\begin{itemize}
 \item Hermiticity
 \begin{align}
  V^{*}_{\lambda'\lambda}(\Delta,P,n)&=V_{\lambda\lambda'}(-\Delta,P,n) 
\,,\nonumber\\ 
A^{*}_{\lambda'\lambda}(\Delta,P,n)&=A_{\lambda\lambda'}(-\Delta,P,n)\,,\nonumber\\
T^{q\,R/L*}_{\lambda'\lambda}(\Delta,P,n)&=-T^{q\,L/R}_{\lambda\lambda'}
(-\Delta , P , n)
\,,\nonumber\\
T^{g\,RR*}_{\lambda'\lambda}(\Delta,P,n)&=T^{g\,LL}_{\lambda\lambda'}
(-\Delta , P , n)\,.
 \end{align}

 \item Light-front parity $\mc{P}_\perp$
 \begin{align}  
V_{\lambda'\lambda}(\Delta,P,n)&=V_{-\lambda'-\lambda}(\widetilde{\Delta},
\widetilde
{P},
\tilde{n}) 
\,,\nonumber\\ 
A_{\lambda'\lambda}(\Delta,P,n)&=-A_{-\lambda'-\lambda}(\widetilde{\Delta},
\widetilde{P}
,
\tilde{n}) \,,\nonumber\\
T^{q\,R/L}_{\lambda'\lambda}(\Delta,P,n)&=-T^{q\,L/R}_{-\lambda'-\lambda}
(\widetilde
{
\Delta},\widetilde{P},
\tilde{n})\,,\nonumber\\
T^{g\,RR}_{\lambda'\lambda}(\Delta,P,n)&=T^{g\,LL}_{-\lambda'-\lambda}
(\widetilde
{
\Delta},\widetilde{P},
\tilde{n})
\,.\label{eq:P}
 \end{align}

 \item Light-front time reversal $\mc{T}_\perp$
 \begin{align}  
V_{\lambda'\lambda}(\Delta,P,n)&=(-1)^{\lambda'-\lambda}\;V_{\lambda\lambda'
}
(-\widetilde { \Delta } , \widetilde { P } ,
\tilde{n}) 
\,,\nonumber\\ 
A_{\lambda'\lambda}(\Delta,P,n)&=(-1)^{\lambda'-\lambda}\;A_{\lambda\lambda'
}
(-\widetilde{\Delta},\widetilde{P},
\tilde{n}) \,,\nonumber\\
T^{q\,R/L}_{\lambda'\lambda}(\Delta,P,n)&=(-1)^{\lambda'-\lambda}\;T^{q\,L/R}_{
\lambda\lambda'}(-\widetilde{
\Delta},\widetilde{P},
\tilde{n})\,,\nonumber\\
T^{g\,RR}_{\lambda'\lambda}(\Delta,P,n)&=(-1)^{\lambda'-\lambda}\;T^{g\,LL}_{
\lambda\lambda'}(-\widetilde{
\Delta},\widetilde{P},
\tilde{n})
\,.
 \end{align}
 \item Finally, $\mc{P}_\perp\mc{T}_\perp$ combined implies
 \begin{align}  
V_{\lambda'\lambda}(\Delta,P,n)&=(-1)^{\lambda'-\lambda}\;V_{
-\lambda-\lambda'}
(-\Delta,P,n) \,,
\nonumber\\ 
A_{\lambda'\lambda}(\Delta,P,n)&=(-1)^{\lambda'-\lambda+1}\;A_{
-\lambda-\lambda'
}
(-\Delta,P,n)\,, \nonumber\\
T^{q\,R/L}_{\lambda'\lambda}(\Delta,P,n)&=(-1)^{\lambda'-\lambda+1}\;T^{q\,R/L}_
{
-\lambda-\lambda'}(-\Delta,P,n)
\,, \nonumber\\
T^{g\,RR/LL}_{\lambda'\lambda}(\Delta,P,n)&=(-1)^{\lambda'-\lambda}\;T^{g\,RR/LL
} _
{
-\lambda-\lambda'}(-\Delta,P,n)
\,.\label{eq:TP}
 \end{align}
 \end{itemize}
where the notation $\widetilde{P}$ is defined in Eq.~(\ref{eq:tilde}).

 


\subsection{Leading twist-2 quark transversity GPDs}
\label{subsec:gpd_t_q}

The leading twist-2 transversity quark GPDs are chiral odd and defined by matrix elements of the tensor correlator
$T^{q\,i}_{\lambda'\lambda}$. They are scalar functions depending on Lorentz invariants $x,\xi,t$ multiplying all possible independent tensor structures that appear in the decomposition of the correlator matrix element.  These tensor structures are built from the available four-vectors $\epsilon,\epsilon',n,P,\Delta$ and the decomposition has to obey the symmetry constraints given in the previous subsection.  We decompose the correlator as
\begin{align} \label{eq:HTq_decomp}
 &\int 
\frac{d\kappa}{2\pi}e^{2ix\kappa(Pn)}\bra{p'\,\lambda'}\bar{\psi}(-\kappa 
n)(i n_\mu\sigma^{\mu i})\psi(\kappa n)\ket{p\,\lambda}=
M\frac{(\epsilon'^{*}n)\epsilon^i-\epsilon'^{*i}(\epsilon n)} {2\sqrt{2}(Pn)}
H^{qT}_1(x,\xi,t)\nonumber\\
&\qquad+M\left[\frac{2P^i(\epsilon n)(\epsilon'^{*} 
n)}{2\sqrt{2}(Pn)^2}-\frac{(\epsilon 
n)\epsilon'^{i*}+\epsilon^i(\epsilon'^{*} n)}{2\sqrt{2}(Pn)}\right] 
H^{qT}_2(x,\xi,t)\nonumber\\
&\qquad+\left[ \frac{(\epsilon'^{*} n)\Delta^i-\epsilon'^{i*}(\Delta n)}{M(Pn)}(\epsilon 
P) - \frac{(\epsilon n)\Delta^i-\epsilon^i (\Delta n)}{M(Pn)}(\epsilon'^{*}P)\right]  H^{qT}_3(x,\xi,t)\nonumber\\
&\qquad+\left[ \frac{(\epsilon'^{*} n)\Delta^i-\epsilon'^{i*}(\Delta n)}{M(Pn)}(\epsilon 
P) + \frac{(\epsilon n)\Delta^i-\epsilon^i (\Delta n)}{M(Pn)}(\epsilon'^{*}P)\right]  H^{qT}_4(x,\xi,t)\nonumber\\
&\qquad+ M\left[ \frac{(\epsilon'^{*} n)\Delta^i-\epsilon'^{i*}(\Delta n)}{2\sqrt{2}(Pn)^2}(\epsilon 
n) + \frac{(\epsilon n)\Delta^i-\epsilon^i (\Delta n)}{2\sqrt{2}(Pn)^2}(\epsilon'^{*}n)\right] 
H^{qT}_5(x,\xi,t)
\nonumber\\
&\qquad+ \frac{(\Delta^i+2\xi P^i)}{M}(\epsilon'^{*} \epsilon)H^{qT}_6(x,\xi,t)
+\frac{(\Delta^i+2\xi P^i)}{M} \frac{(\epsilon'^{*} P)(\epsilon P)}{M^2}
H^{qT}_7(x,\xi,t)\nonumber\\
&\qquad + \left[ \frac{(\epsilon'^{*} n)P^i-\epsilon'^{i*}(Pn)}{M(Pn)}(\epsilon 
P) + \frac{(\epsilon n)P^i-\epsilon^i (Pn)}{M(Pn)}(\epsilon'^{*}P)\right] 
H^{qT}_8(x,\xi,t)\nonumber\\
&\qquad + \left[ \frac{(\epsilon'^{*} n)P^i-\epsilon'^{i*}(Pn)}{M(Pn)}(\epsilon 
P) - \frac{(\epsilon n)P^i-\epsilon^i (Pn)}{M(Pn)}(\epsilon'^{*}P)\right] 
H^{qT}_9(x,\xi,t)\,.
\end{align}

All nine tensor structures are linearly 
independent, consequently so are the nine GPDs. This can be best seen by considering the transformation between the GPDs and helicity amplitudes (see App.~\ref{app:helamps}).  Using the hermiticity, parity and time reversal constraints on the correlators written down in Sec.~\ref{subsec:gpd_rel}, we find the 
following properties of the GPDs:

\begin{itemize}
 \item All nine $H^{qT}_i$ are real.
 \item Even/odd behavior in skewness $\xi$: 
\begin{align}
 &H_i^{qT}(x,-\xi,t)=H_i^{qT}(x,\xi,t) &i\in\{1,4,5,6,7,9\}\,,\nonumber\\
 &H_i^{qT}(x,-\xi,t)=-H_i^{qT}(x,\xi,t) &i\in\{2,3,8\}\,.
\end{align}
\item Sum rules and form factors of local currents:  Due to the odd 
nature of the GPD or the presence of $n^\mu n^\nu/(Pn)^2$ in the accompanying 
tensor, we have the following sum rules that equal zero
\begin{align}\label{eq:q_sums}
 &\int_{-1}^1 \mathrm{d}x\, H^{qT}_i(x,\xi,t)=0  &i\in\{2,3,5,8\}\,.
\end{align}
The first moments of the other 5 GPDs give form factors of local tensor currents.  

\item Forward limit: this corresponds to $\Delta=0,\xi=0, (\epsilon 
P)=(\epsilon'^{*}P)=0$.  The only GPD that does not decouple and is non-zero in this limit is $H^{qT}_1(x,0,0)$. It 
can be connected to the collinear parton distribution function (pdf) $h_1(x)$ 
defined 
in Ref.~\cite{Umnikov:1996hy,Bacchetta:2000jk}:
\begin{equation}
 h_1(x)=H^{qT}_1(x,0,0)\,. 
\end{equation}

\end{itemize}

The correlators of Eq.~(\ref{eq:corr_def}) can be connected to parton-hadron scattering amplitudes in $u$-channel kinematics.  We can thus write the helicity amplitudes of quark-hadron scattering ${\mc{A}^q_{\lambda' \mu';\lambda \mu}}$ [with $\mu$ ($\mu'$) the light-front helicity of the outgoing (incoming) parton line] as certain projections  of Eq.~(\ref{eq:corr_def}) and one has for the chiral odd helicity amplitudes~\cite{Diehl:2001pm}:
\begin{align}\label{eq:hel_def}
 &\mc{A}^q_{\lambda' +;\lambda-}=\frac{1}{2}T^{q\,R}_{\lambda'\lambda} \,,
&\mc{A}^q_{\lambda' 
-;\lambda+}=-\frac{1}{2}T^{q\,L}_{\lambda'\lambda}\,.
\end{align}
Plugging the explicit expressions of the polarization four-vectors of Eq.~(\ref{eq:pol_vectors}) in the decomposition of Eq.~(\ref{eq:HTq_decomp}), we obtain a linear set of transformations between the nine independent helicity amplitudes $\mc{A}^q_{\lambda' +;\lambda-}$ and the nine transversity GPDs $H^{qT}_i$.  This set of equations and their inverse are listed in App.~\ref{app:helamps}.


\subsection{Leading twist-2 gluon transversity GPDs}
\label{subsec:gpd_t_g}
The leading twist-2 transversity gluon GPDs are defined by matrix elements of the tensor correlator
$T^{g\,ij}_{\lambda'\lambda}$. We decompose this correlator as
\begin{align}
 &-\frac{2}{(Pn)}\int 
\frac{d\kappa}{2\pi}e^{2ix\kappa(Pn)}\bra{p'\,\lambda'}\text{Tr}\hat{\bm{S}}
\left
[
 n_\alpha 
G^{\alpha i}(-\kappa n) n_\beta G^{\beta j}(\kappa n) \right]\ket{p\,\lambda}=
\hat{\bm S} \left\{
(\Delta^i+2\xi P^i)\frac{(\epsilon'^{*}n)\epsilon^j-\epsilon'^{j*}(\epsilon 
n)} {(Pn)}
H^{gT}_1(x,\xi,t)\right.\nonumber\\
&\qquad+(\Delta^i+2\xi P^i)\left[\frac{2P^j(\epsilon n)(\epsilon'^{*} 
n)}{(Pn)^2}-\frac{(\epsilon 
n)\epsilon'^{j*}+\epsilon^j(\epsilon'^{*} 
n)}{(Pn)}\right]H^{gT}_2(x,\xi,t)\nonumber\\
&\qquad+\frac{(\Delta^i+2\xi P^i)}{M}\left[ \frac{(\epsilon'^{*} n)\Delta^j-\epsilon'^{j*}(\Delta n)}{M(Pn)}(\epsilon 
P) - \frac{(\epsilon n)\Delta^j-\epsilon^j (\Delta n)}{M(Pn)}(\epsilon'^{*}P)\right] H^{gT}_3(x,\xi,t)\nonumber\\
&\qquad+\frac{(\Delta^i+2\xi P^i)}{M}\left[ \frac{(\epsilon'^{*} n)\Delta^j-\epsilon'^{j*}(\Delta n)}{M(Pn)}(\epsilon 
P) + \frac{(\epsilon n)\Delta^j-\epsilon^j (\Delta n)}{M(Pn)}(\epsilon'^{*}P)\right]  H^{gT}_4(x,\xi,t)\nonumber\\
&\qquad -\left[\frac{(\epsilon'^{*} 
n)P^i-(Pn)\epsilon'^{i*}}{(Pn)}\right]\left[\frac{(\epsilon 
n)P^j-(Pn)\epsilon^j}{(Pn)}\right]H^{gT}_5(x,\xi , t)
\nonumber\\
&\qquad+  \left[\frac{(\epsilon'^{*} 
n)\Delta^i-(\Delta n)\epsilon'^{i*}}{2(Pn)}\right]\left[\frac{(\epsilon 
n)\Delta^j-(\Delta n)\epsilon^j}{2(Pn)}\right]H^{gT}_6(x,\xi,t)
+\frac{(\Delta^i+2\xi P^i)}{M}\frac{(\Delta^j+2\xi P^j)}{M} 
\frac{(\epsilon'^{*} 
P)(\epsilon P)}{M^2}
H^{gT}_7(x,\xi,t)\nonumber\\
&\qquad + \frac{\Delta^i+2\xi P^i}{M}\left[ \frac{(\epsilon'^{*} 
n)P^j-\epsilon'^{j*}(Pn)}{M(Pn)}(\epsilon 
P) + \frac{(\epsilon n)P^j-\epsilon^j (Pn)}{M(Pn)}(\epsilon'^{*}P)\right] 
H^{gT}_8(x,\xi,t)\nonumber\\
&\qquad \left. + \frac{\Delta^i+2\xi P^i}{M} \left[ \frac{(\epsilon'^{*} 
n)P^j-\epsilon'^{j*}(Pn)}{M(Pn)}(\epsilon 
P) - \frac{(\epsilon n)P^j-\epsilon^j (Pn)}{M(Pn)}(\epsilon'^{*}P)\right] 
H^{gT}_9(x,\xi,t)\right\}\,.
\end{align}

The tensor structures that appear in the above equation are linearly 
independent. This is again best observed from the relations between the transversity GPDs and helicity amplitudes written out in App.~\ref{app:helamps}.

 Using the hermiticity, parity and time reversal constraints on the correlators written down in Sec.~\ref{subsec:gpd_rel}, we find the 
following properties of the GPDs:

\begin{itemize}
 \item All nine GPDs are real and even in $x$.
 \item Similarly as for the quark GPDs, the even or odd behavior in skewness $\xi$ is as follows
\begin{align}
 &H_i^{gT}(x,-\xi,t)=H_i^{gT}(x,\xi,t) &i\in\{1,4,5,6,7,9\}\,,\nonumber\\
 &H_i^{gT}(x,-\xi,t)=-H_i^{gT}(x,\xi,t) &i\in\{2,3,8\}\,.
\end{align}
\item Sum rules and form factors of local currents:  Due to the odd 
nature of the GPD, we have the following sum rules that equal zero
\begin{align}
 &\int_{-1}^1 \mathrm{d}x\, H^{gT}_i(x,\xi,t)=0  &i\in\{2,3,8\}\,,
\end{align}
the first moments of the remaining 6 GPDs give form factors of local tensor currents.

\item Forward limit:   The only GPD that does not decouple and is non-zero is $H^{gT}_5(x,0,0)$.
It 
can be connected to structure function $x\Delta$ 
defined in Ref.~\cite{Jaffe:1989xy} [Eq.~(1) within] or the collinear pdf $xh_{1TT}(x)$ 
in Ref.~\cite{Boer:2016xqr} [Eq.~(2.38) within]:
\begin{equation}
 H^{gT}_5(x,0,0)=xh_{1TT}(x)\,.
\end{equation}
This pdf is unique to the spin-1 case as a spin 1/2 hadron cannot 
compensate the gluon helicity flip.
\end{itemize}

The relation between helicity flip gluon-hadron helicity amplitudes $\mc{A}^g_{\lambda' +;\lambda-}$ and the correlators of Eq.~(\ref{eq:corr_def_g}) is given by~\cite{Diehl:2001pm}
\begin{align}\label{eq:hel_def_g}
 &\mc{A}^g_{\lambda' +;\lambda-}=\frac{1}{2}T^{g\,RR}_{\lambda'\lambda} \,,
&\mc{A}^g_{\lambda' -;\lambda+}=\frac{1}{2}T^{g\,LL}_{\lambda'\lambda}\,.
\end{align}
As for the quark sector, we can plug in the explicit expressions for the spin-1 polarization four-vectors and obtain the transformation equations between the helicity amplitudes and the gluon transversity GPDs listed in App.~\ref{app:helamps}.


%
%


\section{Deuteron convolution model: formalism}
\label{sec:d_conv_form}

In this section, we derive the expression of the spin 1 transversity GPDs for 
the 
case of the deuteron in the impulse approximation (IA).  In the IA, we consider 
the 
dominant $NN$ component of the deuteron depicted in the diagram of 
Fig.~\ref{fig:IA}. The two quark lines in the correlators of 
Eqs.~(\ref{eq:corr_def}) are attached to the same nucleon and the second 
nucleon acts as a ``spectator''.  This is a standard first order approximation in the computation of partonic properties of nuclei~\cite{Li:1988rj,Geesaman:1995yd,Kulagin:2004ie,CiofidegliAtti:2007ork,Cosyn:2017fbo,Brodsky:2018zdh}.  The derivation presented here follows the approach used
in Ref.~\cite{Cano:2003ju}: the correlator $T^{qR/L}_{\lambda'\lambda}$ for 
the deuteron is expressed as a convolution of the deuteron light-front wave function with similar correlators for the 
nucleon.  The latter are expressed through the four transversity GPDs of the nucleon.  In the final step the 
correlators can be connected to the transversity deuteron GPDs by inverting the 
relations between the complete set of helicity amplitudes defined by  
Eqs.~(\ref{eq:hel_first}) -- (\ref{eq:hel_last}) and the transversity spin 1 GPDs.

\subsection{Kinematics and conventions}
\label{subsec:conv_kin}

\begin{figure}[htb]
\begin{center}
 \includegraphics[width=0.5\textwidth]{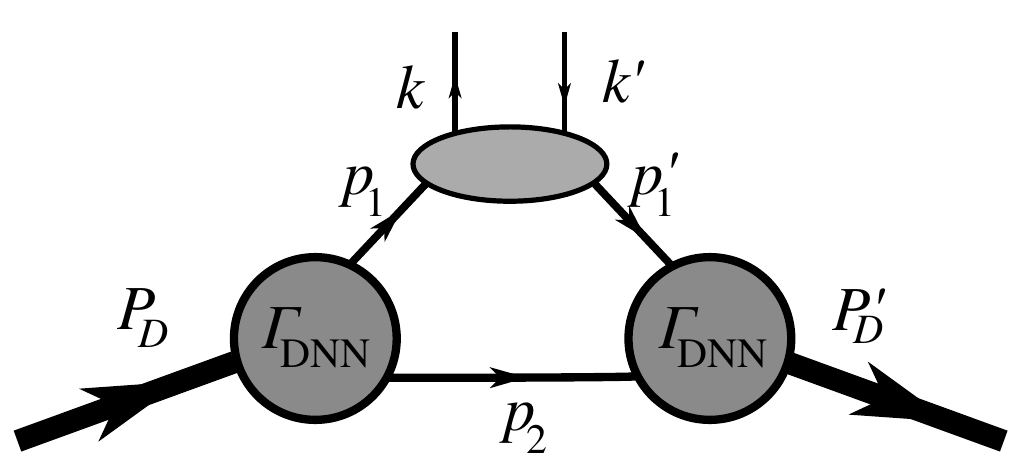}
 \end{center}
 \caption{
 \label{fig:IA}
Diagrammatic depiction of the impulse approximation for the deuteron GPDs, considering the $NN$ component of 
the deuteron, for kinematics where $x>\xi$.}
\end{figure}

As we will be dealing with kinematic variables on both the nuclear and nucleon 
level, we amend the notations of Sec.~\ref{subsec:kin} to differentiate clearly 
between the two.  Using the four-momenta shown in Fig.~\ref{fig:IA}, we introduce the following kinematic variables:
\begin{align}
 &\bar{P}_D=\frac{1}{2}(P_D+P'_D)\,,  &\bar{p}_1=\frac{1}{2}(p_1+p'_1)\,,\nonumber\\
  &\Delta=P'_D-P_D= p'_1-p_1\,,\nonumber\\
  &\xi=-\frac{(\Delta 
n)}{2(\bar{P}_Dn)}\,, &\xi_N=-\frac{(\Delta 
n)}{2(\bar{p}_1n)}\,,\nonumber\\
&\bar{k}=\frac{1}{2}(k+k')\,,\nonumber\\
&x=\frac{(\bar{k}n)}{(\bar{P}_Dn)} \,,
&x_N=\frac{(\bar{k}n)}{(\bar{p}_1n)}\,.
\end{align}
We introduce light-front momentum fractions for the nucleons:
\begin{align}
 &\alpha_1=2\frac{p_1n}{P_Dn} \,,
&\alpha_2=2\frac{p_2n}{P_Dn}=2-\alpha_1\,,\nonumber\\
 &\alpha'_1=2\frac{p'_1n}{P'_Dn} \,,
&\alpha'_2=2\frac{p_2n}{P'_Dn}=2-\alpha'_1\,,\nonumber\\
\end{align}
and we have the following useful identities
\begin{align}
&\frac{(p_1n)}{(\bar{P}_Dn)}=\frac{\alpha_1(1+\xi)}{2} \,,\nonumber\\ 
&\frac{(p'_1n)}{(\bar{P}_Dn)}=\frac{\alpha'_1(1-\xi)}{2}
\,,\nonumber\\ 
&\alpha_1(1+\xi)=\alpha'_1(1-\xi)+4\xi\,,\nonumber\\ 
&\xi_N=\frac{\xi}{\frac{\alpha_1}{2}(1+\xi)-\xi} \,,\nonumber\\ 
&x_N=\frac{x}{\frac{\alpha_1}{2}(1+\xi)-\xi}
\,.
\end{align}
The deuteron light-front wave function depends on the following 
dynamical variable, the three-momentum $\bm{k}_d$ defined by
\begin{align}
 &\frac{k^z_d}{E_{k}}=\alpha_1-1\,,
 &\bm k^\perp_d = \bm p_1^\perp - \frac{\alpha_1}{2}\bm P_D^\perp\,,
 &&E_{k_d}^2 = \bm k_d^2+m^2= \frac{m^2+ (\bm 
k_d^{\perp})^2}{\alpha_1\alpha_2}\,,
 \label{eq:lf_k_mom}
\end{align}
where $m$ is the nucleon mass.  The momentum $\bm{k}_d$ corresponds to the relative momentum of the two on-shell nucleons in the light-front boosted deuteron rest frame~\cite{Frankfurt:1981mk,Keister:1991sb}.
The first two equations follow from the properties of light-front boosts 
while the third equation can be obtained by equating 
$(k_{p}+k_{n})^2=4E_{k_d}^2=2m^2+2(k_{n}k_{p})$, where $k_p,k_n$ are the on-shell nucleon momenta of the intermediate $NN$ state. 

Finally, the phase space element of the active nucleon can be written as
\begin{equation}
 d\Gamma_1 = \frac{dp_1^+d\bm p_1^\perp}{(2\pi)^3 2p_1^+}= \frac{d\alpha_1 d\bm 
p^\perp}{(2\pi)^3 2\alpha_1} = (2-\alpha_1) \frac{d^3 \bm k_d}{(2\pi)^3 2 E_{k_d}}\,.
\end{equation}

\subsection{Deuteron light-front wave function}
\label{subsec:d_lf_wf}

The deuteron light-front wave function~\cite{Frankfurt:1981mk,Kondratyuk:1983kq,Chung:1988my,Keister:1991sb} is given by the overlap of the deuteron 
single-particle state with the on-shell two-nucleon state, where all states are quantized on the light-front:
\begin{equation} \label{eq:lfqm_wf}
  \langle N(p_1,\sigma_1);N(p_2,\sigma_2)|D(P_D,\lambda)
\rangle 
\equiv (2\pi)^\frac{9}{2}2P_D^+\delta(p_1^++p_2^+-P_D^+)\delta(\bm 
p_1^\perp+\bm 
p_2^\perp-\bm P_D^\perp)\Psi_\lambda^D(\bm k_d,\sigma_1,\sigma_2)\,.
\end{equation}
All involved momenta ($P_D,p_1,p_2$) are on their mass shell, which means 
light-front energy (minus component of momentum) is not conserved in the transition $D\to 
NN$.  
For 
the free two-nucleon state in the transition matrix element of 
Eq.~(\ref{eq:lfqm_wf}), an angular momentum decomposition can be performed in 
the 
light-front boosted deuteron rest frame in a way very similar to the case of the 
non-relativistic deuteron wave function.  The relative motion of the two nucleons in the deuteron rest frame 
can be projected on spherical harmonics and for the deuteron a radial $S$-wave 
($l=0$) and $D$-wave ($l=2$) can be coupled to the total spin $S=1$ of the two 
nucleons to obtain total light-front spin $j=1$.
The final form of the deuteron light-front wave function defined through 
Eq.~(\ref{eq:lfqm_wf}) reflects this angular decomposition:
\begin{align}
 \Psi^D_\lambda(\bm 
k,\sigma_1,\sigma_2)&=\sqrt{E_{k}}\sum_{\sigma'_1\sigma'_2}\mathcal{D}^{
\frac{1}{2}}_{\sigma_1\sigma'_1}[R_{fc}(k_{p}^\mu/m)]
\mathcal{D}^{\frac{1}{2}}_{\sigma_2\sigma'_2}[R_{fc}(k_{n}^\mu/m)]
\Phi^D_\lambda(\bm k_d,\sigma'_1,\sigma'_2)\,,
\label{eq:D_lf_wf}
\end{align}
with
\begin{equation}
 \Phi^D_\lambda(\bm 
k_d,\sigma'_1,\sigma'_2)=\sum_{\substack{l=0,2\\\lambda_l\lambda_S}}\langle 
l\lambda_l 
1\lambda_S|1 \lambda\rangle \langle 
\tfrac{1}{2}\sigma'_1\tfrac{1}{2}\sigma'_2|1\lambda_S\rangle 
Y_l^{\lambda_l}(\Omega_{\bm k_d})
\phi_{l}(k)\,,\label{eq:D_nr_wf}
\end{equation}
where the $\phi_{l}(k)$ denote the radial components of the wave 
function and $Y_l^{\lambda_l}(\Omega_{\bm k})$ are the spherical harmonics.

The deuteron light-front wave function has two different features compared to 
the non-relativistic one that deserve highlighting.  First, there is the 
appearance of two Melosh rotations
$\mathcal{D}^{\frac{1}{2}}_{\lambda_i\lambda'_i}[R_{fc}(k_{i}^\mu/m)]$~  
\cite{Melosh:1974cu} in Eq.~(\ref{eq:D_lf_wf}) that encode relativistic spin effects arising from the quantization of particle states (and spin) on the light-front. Second, the dynamical variable 
that appears in the light-front wave function is 
the three-momentum $\bm k$.  In the calculations presented in this article the radial wave functions $\phi_{l}(k)$ are identified with those from non-relativistic wave function parameterizations.  We want to stress that this does not correspond with approximating the
light-front wave function with the non-relativistic one given the differences pointed out above.  This approach can be justified for momenta up to a few 100 MeV given the small binding energy of the deuteron.  In Ref.~\cite{Miller:2009fc}, an explicit comparison between the instant-form and front-form wave
function for a two-particle bound state was carried out in a toy model. The connection
between the non-relativistic instant form and light-front wave function as in Eq.~(\ref{eq:D_lf_wf})  was
found to hold for $\epsilon_B/M_D < 0.002$ (with $\epsilon_B, M_D$ the deuteron binding energy and mass), which holds for the deuteron case.

\subsection{Nucleon chiral odd quark GPDs}
\label{subsec:nucl_gpd}
As the tensor correlator defining the nucleon chiral odd quark GPDs appears in the 
IA derivation, we briefly summarize expressions for these in this subsection.
We use the standard parametrization for the nucleon chiral odd quark GPDs 
introduced in Ref.~\cite{Diehl:2001pm}:
\begin{multline}
 \int 
\frac{d\kappa}{2\pi}e^{2ix_N\kappa(\bar{p}_1n)}\bra{p'_1\,\sigma'_1}\bar{\psi}
(-\kappa 
n)(i n_\mu\sigma^{\mu i})\psi(\kappa 
n)\ket{p_1\,\sigma_1}=
\frac{1}{2(\bar{p}_1 n)}\bar{u}(p'_1,\sigma'_1)\left[ 
H^q_T(i n_\mu \sigma^{\mu i}) + \widetilde{H}^q_T 
\frac{(\bar{p}_1n)\Delta^i-(\Delta n)\bar{p}_1^i}{m^2}\right.\\
\left.
+E^q_T \frac{(\gamma 
n)\Delta^i-(\Delta n)\gamma^i}{2m} +\widetilde{E}^q_T \frac{(\gamma 
n)\bar{p}_1^i-(\bar{p}_1 n)\gamma^i}{m}
\right]u(p_1,\sigma_1)\,.\label{eq:nucl_odd}
\end{multline}
Substituting the standard light-front spinors~\cite{Brodsky:1997de}, we list explicit expressions for 
the spinor bilinears multiplying the GPDs in the above expression.  For $\pm$ 
and $\mp$ appearing in the following expressions the upper sign comes with the 
$R$ component, the lower one with the $L$ component.  We have
\begin{align}
 \frac{1}{2(\bar{p}_1 n)}\bar{u}(p'_1,\sigma'_1) (i n_\mu \sigma^{\mu R/L})
u(p_1,\sigma_1)&=
-\delta_{-\sigma'_1,\sigma_1} (2\sigma_1 \mp 
1)\sqrt{1-\xi_N^2}\,,\nonumber\\
\frac{1}{2(\bar{p}_1 n)}\bar{u}(p'_1,\sigma'_1) \left[
\frac{(\bar{p}_1n)\Delta^{R/L}-(\Delta 
n)\bar{p}_1^{R/L}}{m^2} \right] u(p_1,\sigma_1)&=
\delta_{\sigma'_1,\sigma}\frac{\sqrt{t_{0N}-t}}{m}e^{\pm 
i\phi_1}-\delta_{-\sigma'_1,\sigma_1}\sigma_1\frac{\sqrt{1-\xi_N^2}(t_{0N}-t)}{
m^2}
e^{ (2\sigma_1\pm1)i\phi_1}\,, \nonumber\\
\frac{1}{2(\bar{p}_1 n)}\bar{u}(p'_1,\sigma'_1) \left[
\frac{(\gamma 
n)\Delta^{R/L}-(\Delta n)\gamma^{R/L}}{2m} \right] u(p_1,\sigma_1)&=
\delta_{\sigma'_1,\sigma_1}\frac{(1\mp 2\sigma_1 
\xi_N)\sqrt{t_{0N}-t}}{2m}e^{\pm 
i\phi_1} + \delta_{-\sigma'_1,\sigma_1} (2\sigma_1\mp 
1)\frac{\xi_N^2}{\sqrt{1-\xi_N^2}}\,,\nonumber\\
\frac{1}{2(\bar{p}_1 n)}\bar{u}(p'_1,\sigma'_1) \left[
\frac{(\gamma 
n)\bar{p}_1^{R/L}-(\bar{p}_1 n)\gamma^{R/L}}{m} \right] u(p_1,\sigma_1)&=
\pm \delta_{\sigma'_1,\sigma_1}2\sigma_1\frac{(1\mp 2\sigma_1 
\xi_N)\sqrt{t_{0N}-t}}{2m}e^{\pm 
i\phi_1} - \delta_{-\sigma'_1,\sigma_1} (2\sigma_1\mp 
1)\frac{\xi_N}{\sqrt{1-\xi_N^2}}\,, \label{eq:spinor_expr}
\end{align}
where $\phi_1$ is the azimuthal angle of the four-vector $\Delta + 2\xi_N 
\bar{p}_1$ and
\begin{equation}
 t_{0N}=-\frac{4m^2\xi_N^2}{1-\xi_N^2}\,.
\end{equation}

\subsection{Impulse approximation derivation}
\label{subsec:conv_derivation}

As the following derivation does not depend on the exact operator in the 
correlator, we leave it unspecified and 
call it $\hat{A}$.  Consequently the equations below apply to any 
quark-quark or gluon-gluon GPD correlator written down in Subsec.~\ref{subsec:gpd_rel}.

We start by inserting two complete sets of on-shell two-nucleon states in the 
correlator, use Eq.~(\ref{eq:lfqm_wf}) to introduce the deuteron light-front 
wave functions and Eq.~(\ref{eq:sp_norm}) to evaluate the integrations over the spectator nucleon phase space elements:
\begin{multline}\label{eq:IA_deriv}
 \int 
\frac{d\kappa}{2\pi}e^{2ix\kappa(\bar{P}_Dn)}\bra{P'_D\,\lambda'}\hat{A}\ket{
P_D\ ,\lambda}=\sum_N\int \frac{dp^+_1d\bm 
p^\perp_1}{2p_1^+}
\frac{dp'^{+}_1d\bm p'^{\perp}_1}{2p'^{+}_1}\frac{dp^+_2d\bm 
p^\perp_2}{2p_2^+}2P^{+}_D\,2P'^{+}_D
\delta^{+\perp}(P'_D-p'_1-p_2)
\\
\times \delta^{
+\perp}(P_D-p_1-p_2)
\Theta(\tfrac{\alpha_1}{2}(1+\xi)-|x|-\xi)\left[\Theta(\xi)\Theta(\tfrac{
\alpha_1 } { 2 }
(1+\xi)-2\xi)+\Theta(-\xi)\Theta(\tfrac{\alpha_1}{2}(1+\xi))\right]\\
\times \sum_{\sigma_1\sigma'_1\sigma_2}\Psi_{\lambda'}^{*D}(\bm 
k'_d,\sigma'_1,\sigma_2)\Psi_\lambda^{D}(\bm 
k_d,\sigma_1,\sigma_2)\int 
\frac{d\kappa}{2\pi}e^{2ix_N\kappa(\bar{p}_1n)}\bra{p_1'\,\sigma'_1}\hat{A}\ket{
p_1\,\sigma_1}\\
=\sum_N\int \frac{d\alpha_1d\bm 
p^\perp_1}{\alpha_1}
\frac{d\alpha'_1d\bm 
p'^{\perp}_1}{\alpha'_1} 
\frac{P_D^+P'^{+}_D}{2p_2^+}\delta(-\Delta^+-p^{+}_1+p'^{+} _1)\delta(-\bm 
\Delta^\perp -\bm p^\perp_1+\bm 
p'^{\perp}_1)\Theta(\tfrac{\alpha_1}{2}(1+\xi)-|x|-\xi)\\\times 
\left[\Theta(\xi)\Theta(\tfrac{
\alpha_1 } { 2
}
(1+\xi)-2\xi)+\Theta(-\xi)\Theta(\tfrac{\alpha_1}{2}(1+\xi))\right]\sum_{
\sigma_1\sigma'_1\sigma_2}\Psi_{\lambda'}^{*D}
(\bm 
k'_d,\sigma'_1,\sigma_2)\Psi_\lambda^{D}(\bm 
k_d,\sigma_1,\sigma_2)\int 
\frac{d\kappa}{2\pi}e^{2ix_N\kappa(\bar{p}_1n)}\bra{p_1'\,\sigma'_1}\hat{A}\ket{
p_1\,\sigma_1}\\
=\sum_N\int \frac{d\alpha_1d\bm 
p^\perp_1}{\alpha_1}
\frac{d\alpha'_1d\bm 
p'^{\perp}_1}{\alpha'_1} 
\frac{2}{2-\alpha_1}\delta(\alpha'_1-\alpha_1\frac{1+\xi}{1-\xi}
+4\frac{\xi}{1-\xi})\delta(-\bm 
\Delta^\perp -\bm p^\perp_1+\bm 
p'^{\perp}_1)\Theta(\tfrac{\alpha_1}{2}(1+\xi)-|x|-\xi)\\\times 
\left[\Theta(\xi)\Theta(\tfrac{
\alpha_1 } { 2
}
(1+\xi)-2\xi)+\Theta(-\xi)\Theta(\tfrac{\alpha_1}{2}(1+\xi))\right]\sum_{
\sigma_1\sigma'_1\sigma_2}\Psi_{\lambda'}^{*D}
(\bm 
k'_d,\sigma'_1,\sigma_2)\Psi_\lambda^{D}(\bm 
k_d,\sigma_1,\sigma_2)\int 
\frac{d\kappa}{2\pi}e^{2ix_N\kappa(\bar{p}_1n)}\bra{p_1'\,\sigma'_1}\hat{A}\ket{
p_1\,\sigma_1}\\
=2\sum_N\int \frac{d\alpha_1d\bm 
k_d^{\perp}}{\alpha_1(2-\alpha_1)} \frac{d\alpha'_1d\bm 
{k'}^{\perp}_d}{\alpha'_1} \delta(\alpha'_1-\alpha_1\frac{1+\xi}{1-\xi}
+4\frac{\xi}{1-\xi})\delta(\bm 
{k'}^{\perp}_d-\bm k_d^\perp-\frac{1-\tfrac{\alpha_1}{2}}{ 1-\xi
}\bm \Delta^\perp-2\xi\frac{1-\tfrac{\alpha_1}{2}}{1-\xi}\bar{\bm P}_D^\perp 
)\Theta(\tfrac{\alpha_1}{2}(1+\xi)-|x|-\xi)\\
\times\left[\Theta(\xi)\Theta(\tfrac{
\alpha_1 } { 2
}
(1+\xi)-2\xi)+\Theta(-\xi)\Theta(\tfrac{\alpha_1}{2}(1+\xi))\right]\sum_{
\sigma_1\sigma'_1\sigma_2}\Psi_{\lambda'}^{*D}(\bm 
k'_d,\sigma'_1,\sigma_2)\Psi_\lambda^{D}(\bm 
k_d,\sigma_1,\sigma_2)\int 
\frac{d\kappa}{2\pi}e^{2ix_N\kappa(\bar{p}_1n)}\bra{p_1'\,\sigma'_1}\hat{A}\ket{
p_1\,\sigma_1}\,.
\end{multline}
In the third step a factor $2\bar{P}_D$ was brought into the Dirac delta 
function for the plus components.  The sum $N$ is over the two possible active 
nucleons.  The Heaviside functions originate from the requirement of positive 
light-front plus components for the on-shell intermediate states. The $|x|>|\xi|$ region gives the first 
Heaviside, the ERBL region the remaining ones.  

Up to here the derivation is valid for any correlator considered in Subec.~\ref{subsec:gpd_rel}.
In the next step, we specialize to the case of the twist-2 chiral odd quark GPDs. By 
taking 
$\hat{A}=\bar{\psi}(-\kappa 
n)(in_\mu\sigma^{\mu R/L})\psi(\kappa 
n)$ in Eq.~(\ref{eq:IA_deriv}) and using 
Eqs.~(\ref{eq:nucl_odd}) and (\ref{eq:spinor_expr}) we arrive at
\begin{multline}\label{eq:conv_final}
 T^{R/L}_{\lambda'\lambda} = 4\int \frac{d\alpha_1d\bm 
k^{\perp}_d}{\alpha_1(2-\alpha_1)}\frac{d\alpha'_1d\bm 
k'^{\perp}_d}{\alpha'_1} \delta(\alpha'_1-\alpha_1\frac{1+\xi}{1-\xi}
+4\frac{\xi}{1-\xi})\delta(\bm 
k'^{\perp}_d-\bm k^\perp_d-\frac{1-\tfrac{\alpha_1}{2}}{ 1-\xi
}\bm \Delta^\perp-2\xi\frac{1-\tfrac{\alpha_1}{2}}{1-\xi}\bar{\bm P}_D^\perp 
)\\
\times\Theta(\tfrac{\alpha_1}{2}(1+\xi)-|x|-\xi)
\left[\Theta(\xi) \Theta(\tfrac{\alpha_1}{2
}
(1+\xi)-2\xi) + \Theta(-\xi) \Theta(\tfrac{\alpha_1}{2
}
(1+\xi))\right]
\\
\sum_{\sigma_1\sigma'_1\sigma_2}\Psi_{\lambda'}^{*D}(\bm 
k'_d,\sigma'_1,\sigma_2)\Psi_\lambda^{D}(\bm 
k_d,\sigma_1,\sigma_2)\,. \left[ -\delta_{-\sigma'_1,\sigma_1} (2\sigma_1 \mp 
1)\sqrt{1-\xi_N^2} H^{\text{IS}}_T(x_N,\xi_N,t)\right. \\
\left. +
\left(\delta_{\sigma'_1,\sigma_1}\frac{\sqrt{t_0-t}}{m}e^{\pm 
i\phi_1}-\delta_{-\sigma'_1,\sigma_1}\sigma_1\frac{\sqrt{1-\xi_N^2}(t_0-t)}{m^2}
e^{ (2\sigma_1\pm1)i\phi_1}\right) \widetilde{H}^{\text{IS}}_T(x_N,\xi_N,t)
\right.\\
\left. +\left(\delta_{\sigma'_1,\sigma_1}\frac{(1\mp 2\sigma_1 
\xi_N)\sqrt{t_0-t}}{2m}e^{\pm 
i\phi_1} + \delta_{-\sigma'_1,\sigma_1} (2\sigma_1\mp 
1)\frac{\xi_N^2}{\sqrt{1-\xi_N^2}} \right) 
E^{\text{IS}}_T(x_N,\xi_N,t)\right.\\
\left.+ \left( \pm \delta_{\sigma'_1,\sigma_1}2\sigma_1\frac{(1\mp 2\sigma_1 
\xi_N)\sqrt{t_0-t}}{2m}e^{\pm 
i\phi_1} - \delta_{-\sigma'_1,\sigma_1} (2\sigma_1\mp 
1)\frac{\xi_N}{\sqrt{1-\xi_N^2}}\right)\widetilde{E}^{\text{IS}}
_T(x_N,\xi_N,t)
\right]\,,
\end{multline}
where the nucleon GPDs are the isoscalar combinations
\begin{equation} 
X^{\text{IS}}(x_N,\xi_N,t)=\frac{1}{2}\left(X^{u}(x_N,\xi_N,t)+X^{d}(x_N,\xi_N,
t) \right)\,,
\end{equation}
originating from the isoscalar nature of the deuteron $np$ component considered here.
Because of the non-conservation of the minus component in the $D\rightarrow NN$ vertex, the $t$ appearing in 
the nucleon GPDs is in principle different from the $t$ defined in the 
beginning [i.e. for the deuteron as defined in Eq.~(\ref{eq:kin_d})].  Due to the small binding energy 
$\epsilon_B$ of the deuteron, the difference between the two will go as $\epsilon_B$ over 
some larger scale and can be neglected in a first approximation.
The deuteron transversity GPDs can be obtained from Eq.~(\ref{eq:conv_final}) by first calculating the helicity amplitudes [Eq.~(\ref{eq:hel_def})] and subsequently using the results of App.~C [Eqs.~(\ref{eq:ampstoGPD_first}) -- (\ref{eq:ampstoGPD_last})] to compute the GPDs from the helicity amplitudes.

Comparing our derivation with the one presented in Ref.~\cite{Cano:2003ju}, we notice the 
following differences.  Equation numbers mentioned below refer to 
the ones in Ref.~\cite{Cano:2003ju}:
\begin{itemize}
 \item Eq.~(A2) is missing a factor 1/2 in the right-hand side so that the deuteron particle state is correctly normalized.  As a consequence, 
 Eq.~(19) (and following) need an additional factor 1/4.
 \item Eq.~(29) should have a prefactor of $\frac{1}{16\pi^3}$.  There is a factor of 2 missing in the transition from Eq.~(28) to (29) and a factor of 1/4 from the first bullet above.
 \item The phase of Eq.~(31) should read $\eta_\lambda=(2\lambda\tilde{\Delta}_x+i\tilde{\Delta}_y)/|\tilde{\Delta}_\perp|$.  This can also be inferred from the helicity 
amplitudes written down in Eq.~(61) of Ref.~\cite{Diehl:2003ny}, where a factor 
$e^{\pm i\phi}$ is written in the $\mc{A}_{\mp+,\pm+}$ amplitudes.  
\end{itemize}


\section{Deuteron convolution model: results}
\label{sec:results}
In this section, we use Eq.~(\ref{eq:conv_final}) in combination with Eq.~(\ref{eq:hel_def}) and Eqs.~(\ref{eq:ampstoGPD_first}) -- (\ref{eq:ampstoGPD_last}) to compute the helicity amplitudes and transversity GPDs in the quark sector for the deuteron. For the chiral odd nucleon GPDs, we use  the 
parametrization of Goloskokov and Kroll (GK)~\cite{Goloskokov:2011rd}, evaluated at a scale of $\mu=2~$GeV, and implemented 
in three different models in Ref.~\cite{Pire:2017lfj} (the figures below use 
``model 2'' therein, which has $\tilde{H}_T=H_T$, $E_T=\bar{E}_T-2H_T$, $\tilde{E}_T=0$ ).  In the forward limit of this parametrization the helicity pdfs enter~\cite{Beiyad:2010cxa}. For these we use the 
parametrization of Ref.~\cite{Martin:2009iq}.  We use the AV18 parametrization 
of the deuteron wave function~\cite{Wiringa:1994wb} unless otherwise noted.  As from this section on we are only dealing with quark helicity amplitudes and GPDs, we omit the superscript $q$ for those quantities.  We verified that the computed deuteron helicity amplitudes obey all the symmetry constraints listed in Subsec.~\ref{subsec:gpd_rel}, up to the numerical accuracy imposed on the integrations over the active nucleon phase space.

\begin{figure}[htb]
\begin{center}
 \includegraphics[width=0.8\textwidth]{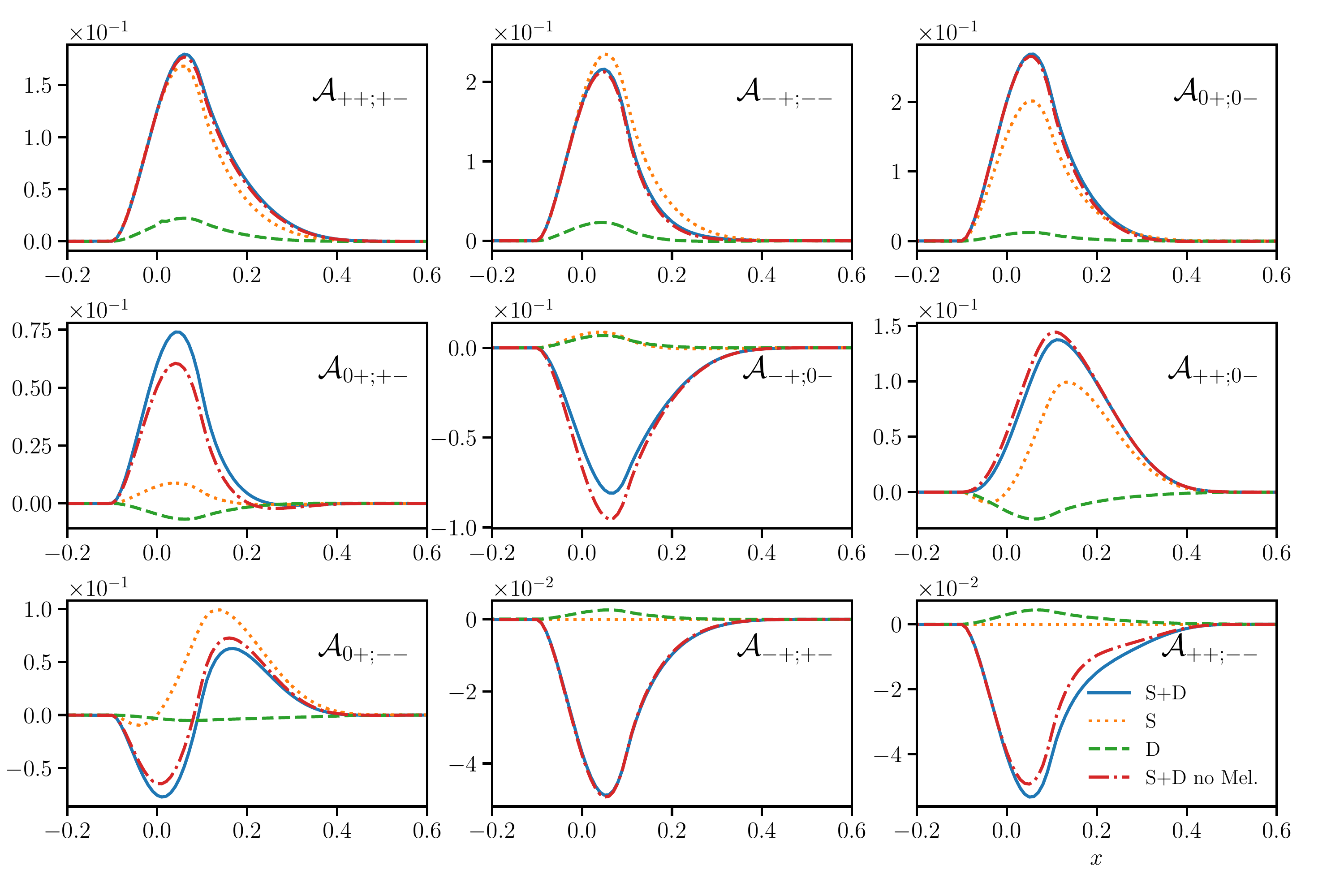}
 \end{center}
 \caption{
 \label{fig:amps}
(Color online) Deuteron quark helicity amplitudes computed in the convolution formalism, at 
$\xi=0.1$,$t=-0.25$~GeV$^2$.  Full blue curve includes the full deuteron wave 
function, dotted orange (dashed green) only includes the deuteron 
radial $S$-($D$-)wave and the dashed-dotted red corve omits the Melosh rotations in 
the light-front deuteron wave function.}
\end{figure}

\begin{figure}[htb]
\begin{center}
 \includegraphics[width=0.8\textwidth]{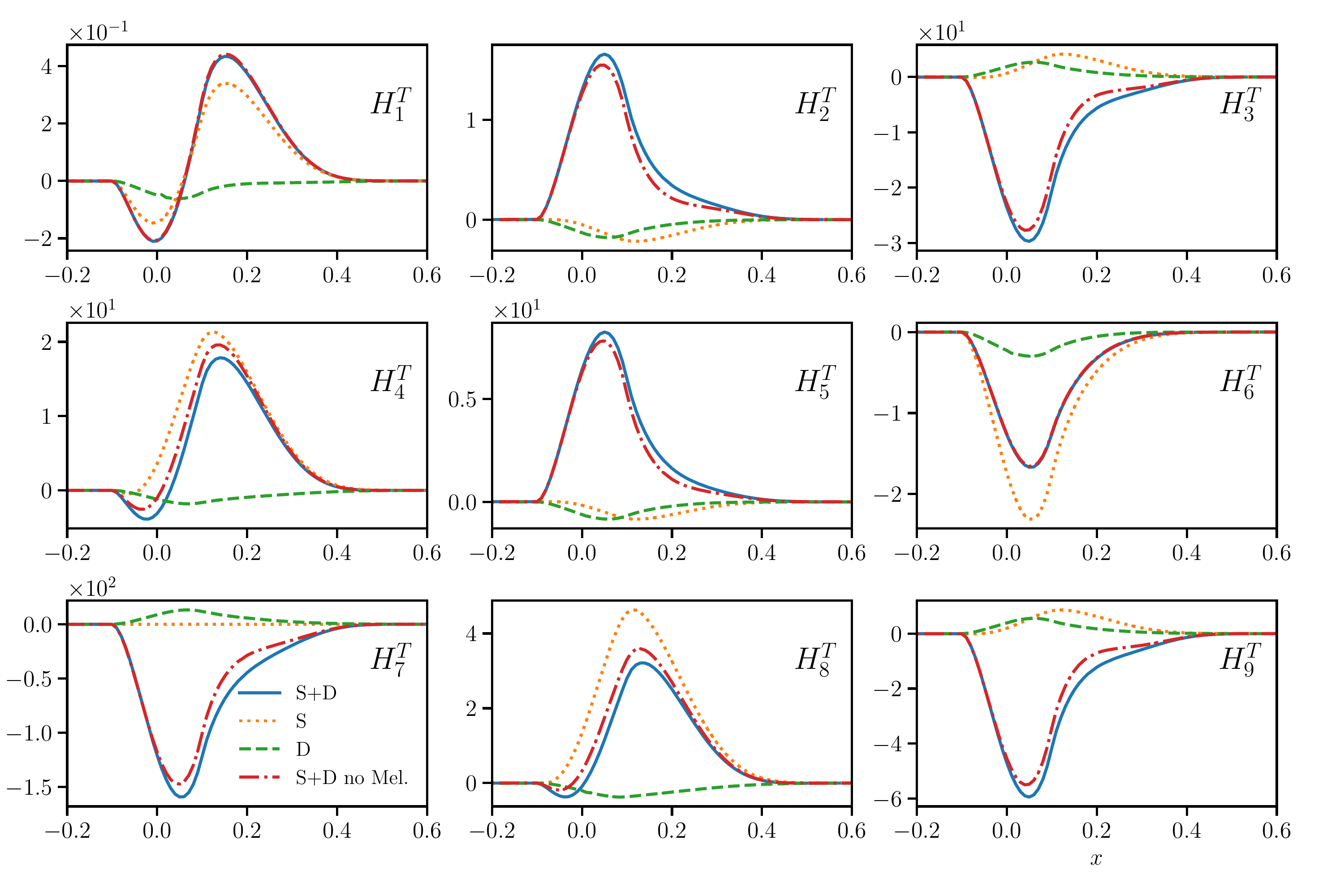}
 \end{center}
 \caption{
 \label{fig:gpds}
(Color online) Deuteron quark transversity GPDs computed in the convolution formalism, at 
$\xi=0.1$,$t=-0.25$~GeV$^2$.  Curves as in Fig.~\ref{fig:amps}}.
\end{figure}

Figures~\ref{fig:amps} and \ref{fig:gpds} show the helicity amplitudes and 
quark transversity GPDs of the deuteron as a function of $x$ (where $-1\leq x \leq 1$) , in kinematics 
$\xi=0.1$ and $t=-0.25$~GeV$^2$. 
Next to the total result, Figs.~\ref{fig:amps} and \ref{fig:gpds} also show the 
separate contributions to 
the helicity amplitudes and GPDs when only including the deuteron radial $S$- or 
$D$-wave.  The remaining difference with the total result originates from 
$S$-$D$ interference contributions.  For the helicity amplitudes, 
one observes that the deuteron helicity conserving ones (top row of Fig.~\ref{fig:amps}) are dominated by the pure $S$-wave contribution, whereas the ones with 
a helicity change for the deuteron receive sizeable contributions from $S$-$D$ 
interference terms.  The effect of the Melosh rotations is generally smallest 
in amplitudes dominated by the $S$-wave contribution.  Lastly, the two 
amplitudes with a complete deuteron 
helicity flip (Fig.~\ref{fig:amps}  bottom row, middle and right panel) are 
identically zero for the $S$-wave as there is no orbital angular momentum 
available in the deuteron to compensate the change in helicities (two units for 
the deuteron, one unit for the quark).  

 \begin{figure}[htb]
 \begin{center}
  \includegraphics[width=0.8\textwidth]{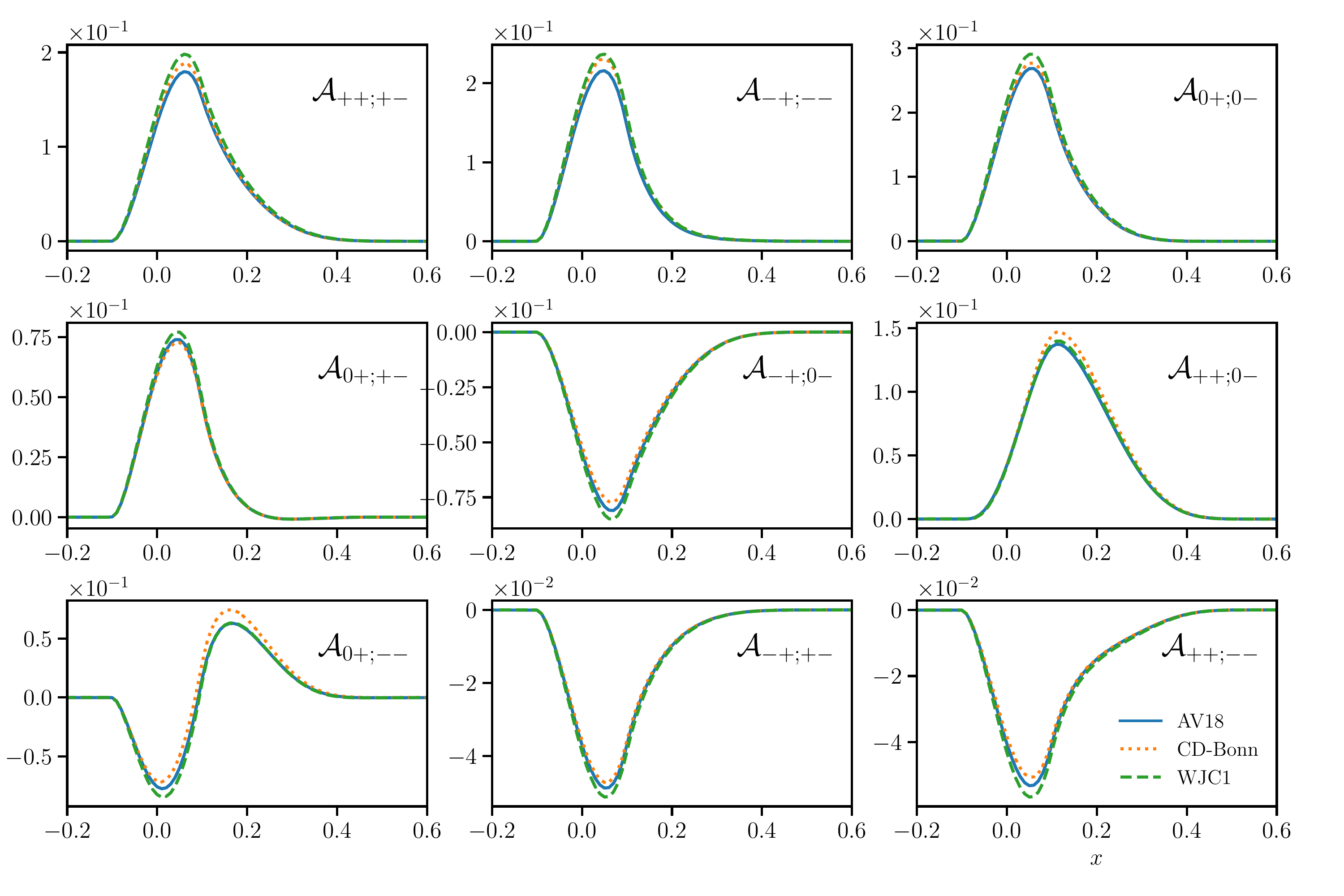}
  \end{center}
  \caption{
  \label{fig:amps_wf}
(Color online)  Deuteron quark helicity amplitudes computed in the convolution formalism, at 
 $\xi=0.1$,$t=-0.25$~GeV$^2$ with different deuteron wave functions.  Deuteron wave functions are CD-Bonn~\cite{Machleidt:2000ge},
 WJC-1~\cite{Gross:2008ps} and AV18~\cite{Wiringa:1994wb}. }
 \end{figure}
 
 \begin{figure}[htb]
 \begin{center}
  \includegraphics[width=0.8\textwidth]{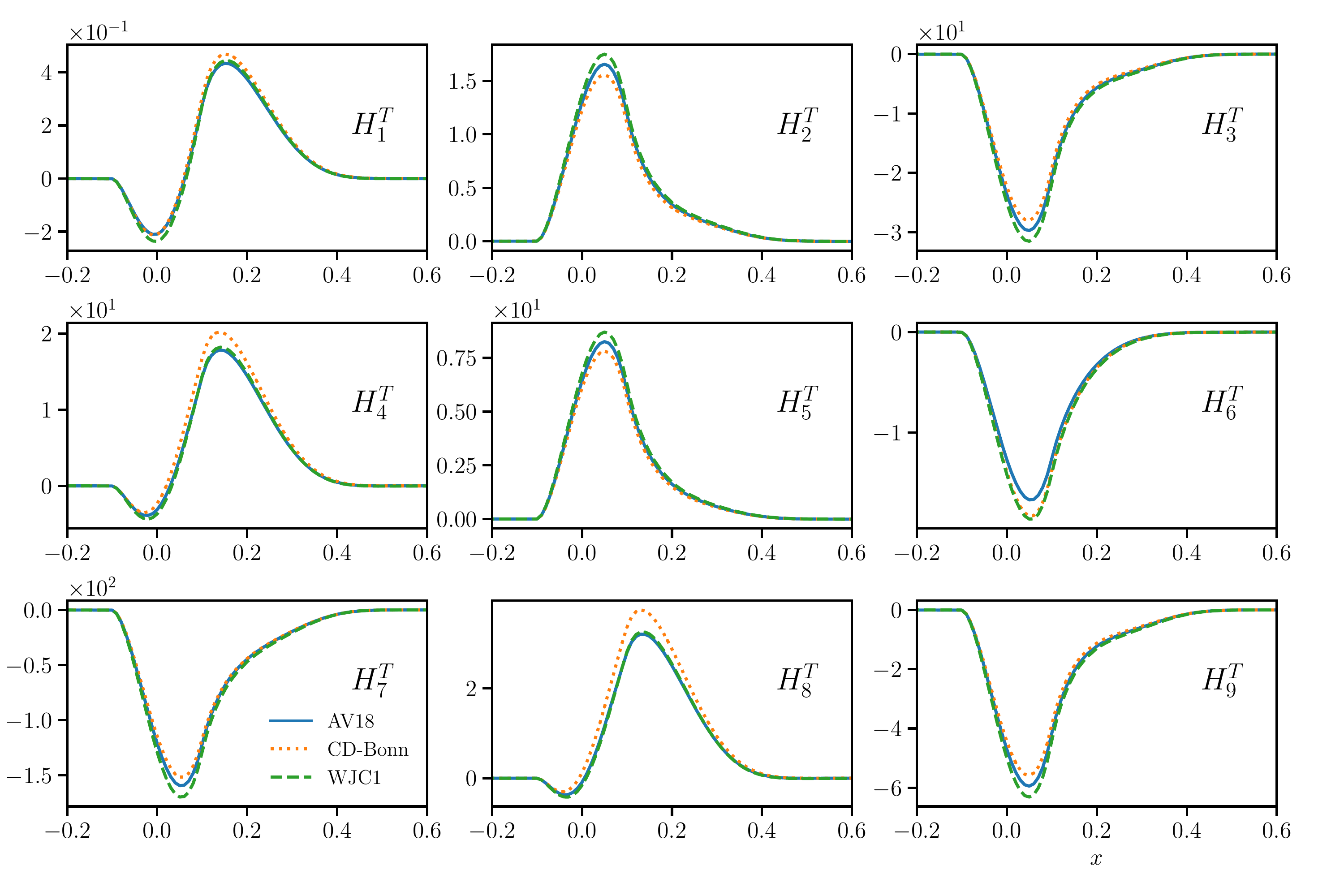}
  \end{center}
  \caption{
  \label{fig:gpds_wf}
(Color online)  Deuteron quark transversity GPDs computed in the convolution formalism, at 
 $\xi=0.1$,$t=-0.25$~GeV$^2$. Comparison between different wave functions.  
}
 \end{figure}

We compared calculations with the three slightly 
different implementations of the chiral odd nucleon GPD models used in Ref.~\cite{Pire:2017lfj}. The results proved to be rather insensitive to these choices as changes in the final deuteron GPDs were in the order of a few 
percent maximum.
 Similary, Figs.~\ref{fig:amps_wf} and 
 \ref{fig:gpds_wf} show the use of three different deuteron wave functions in the calculation: the CD-Bonn~\cite{Machleidt:2000ge} has a soft high-momentum tail, the WJC-1~\cite{Gross:2008ps} a hard one, and the AV18~\cite{Wiringa:1994wb} wave function is in-between.  Consequently, the differences between the different parametrizations included here are largest at high $\bm p^\perp$ or $\alpha_1$ close to its lower (0) and upper (2) bound.
Both the amplitudes and GPDs are in general rather insensitive to the wave 
 function details, even for the amplitudes that do not receive a pure $S$-wave 
 contribution, and which are dominated by high relative $NN$ momenta in the convolution.

\begin{figure}[htb]
\begin{center}
 \includegraphics[width=0.8\textwidth]{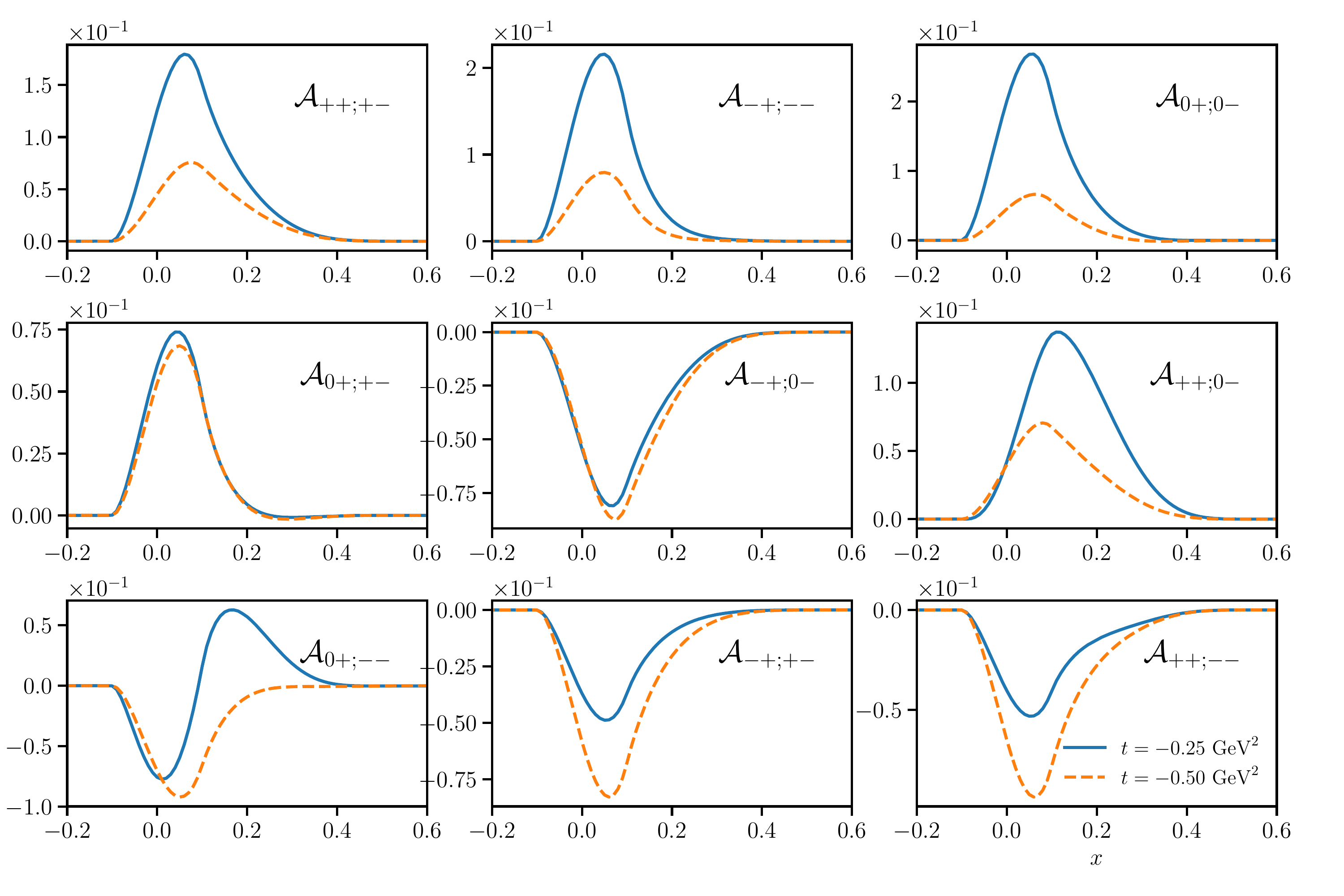}
 \end{center}
 \caption{
 \label{fig:amps_t50}
(Color online) Deuteron quark helicity amplitudes computed in the convolution formalism, at 
$\xi=0.1$ and two values of momentum transfer $t$.}
\end{figure}

\begin{figure}[htb]
\begin{center}
 \includegraphics[width=0.8\textwidth]{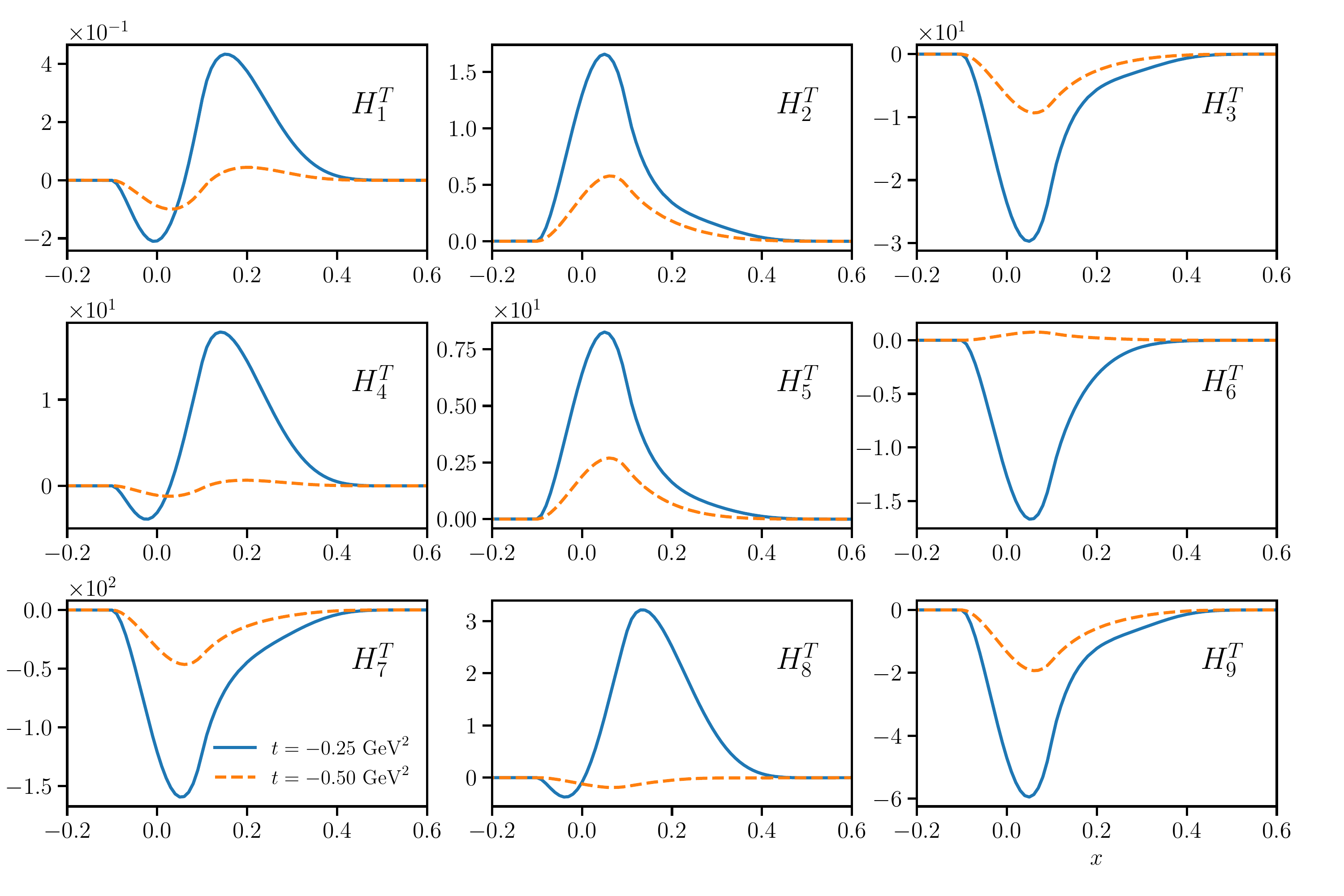}
 \end{center}
 \caption{
 \label{fig:gpds_t50}
(Color online) Deuteron quark transversity GPDs computed in the convolution formalism, at 
$\xi=0.1$ and two values of momentum transfer $t$.}
\end{figure}

Figures~\ref{fig:amps_t50} and \ref{fig:gpds_t50} show calculations at two values 
of momentum transfer.  Helicity amplitudes without deuteron helicity flip 
shrink in size with higher momentum transfer. The amplitudes with a single 
helicity flip also become slightly smaller but the effect is not as large.  
Finally, 
the amplitudes with a double helicity flip grow in size.  This reflects the 
role 
angular momentum plays in these amplitudes, being supplied by the momentum 
transfer.  The GPDs are in general smaller at higher momentum transfer. 
$H^T_6$ has flipped sign, this is caused by the fact that $H^T_6$ is 
proportional to the sum of helicity conserving and double helicity flip 
amplitudes (entering with different sign) [Eqs.~(\ref{eq:HT6}) and ~(\ref{eq:HT8})].

\begin{figure}[htb]
\begin{center}
 \includegraphics[width=0.8\textwidth]{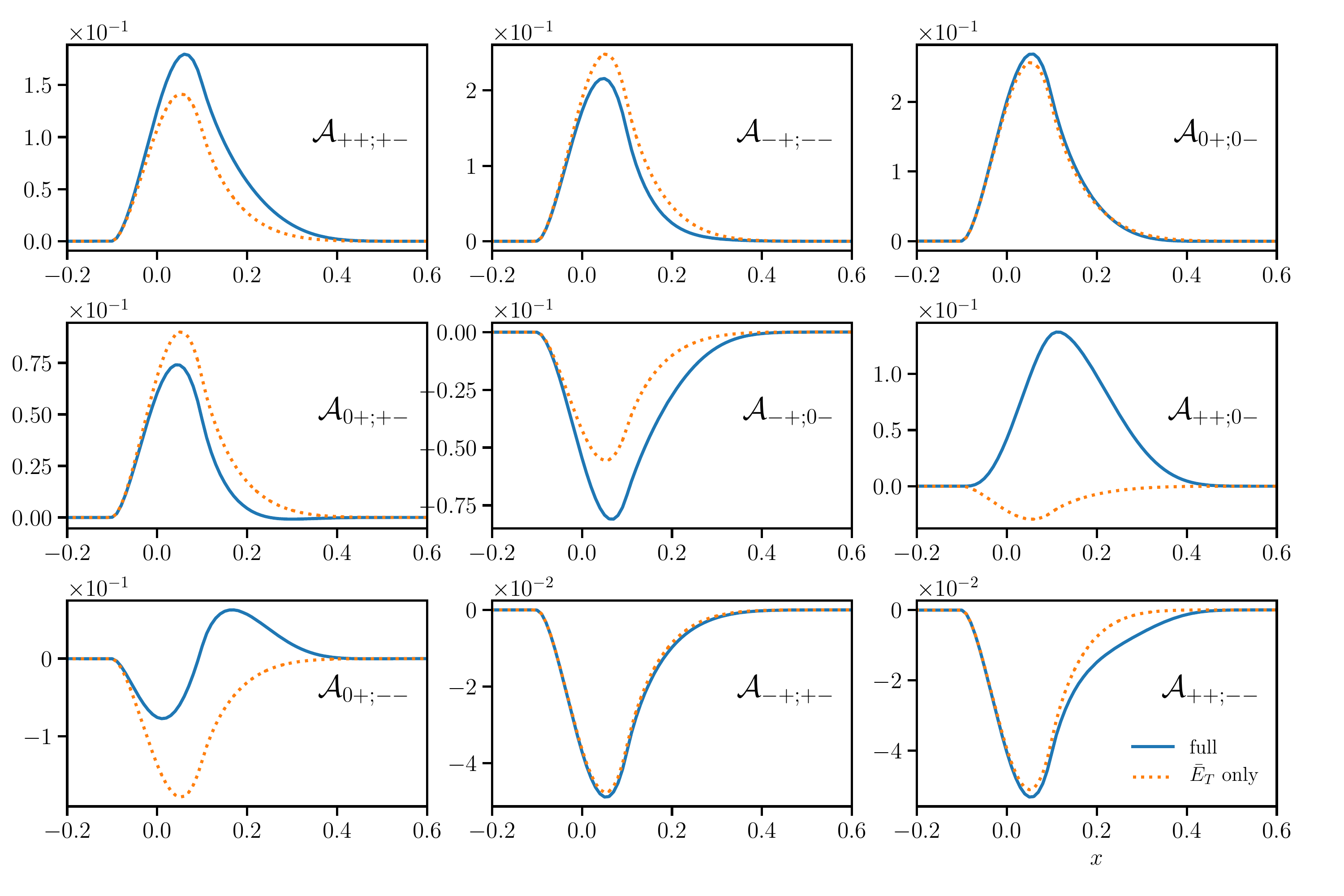}
 \end{center}
 \caption{
 \label{fig:amps_ht}
(Color online) Deuteron quark helicity amplitudes computed in the convolution formalism, at 
$\xi=0.1$,$t=-0.25$~GeV$^2$.  Dashed curve is a calculation only considering 
the $\bar{E}_T$ nucleon GPD.}
\label{fig:amp_noht}
\end{figure}

\begin{figure}[htb]
\begin{center}
 \includegraphics[width=0.8\textwidth]{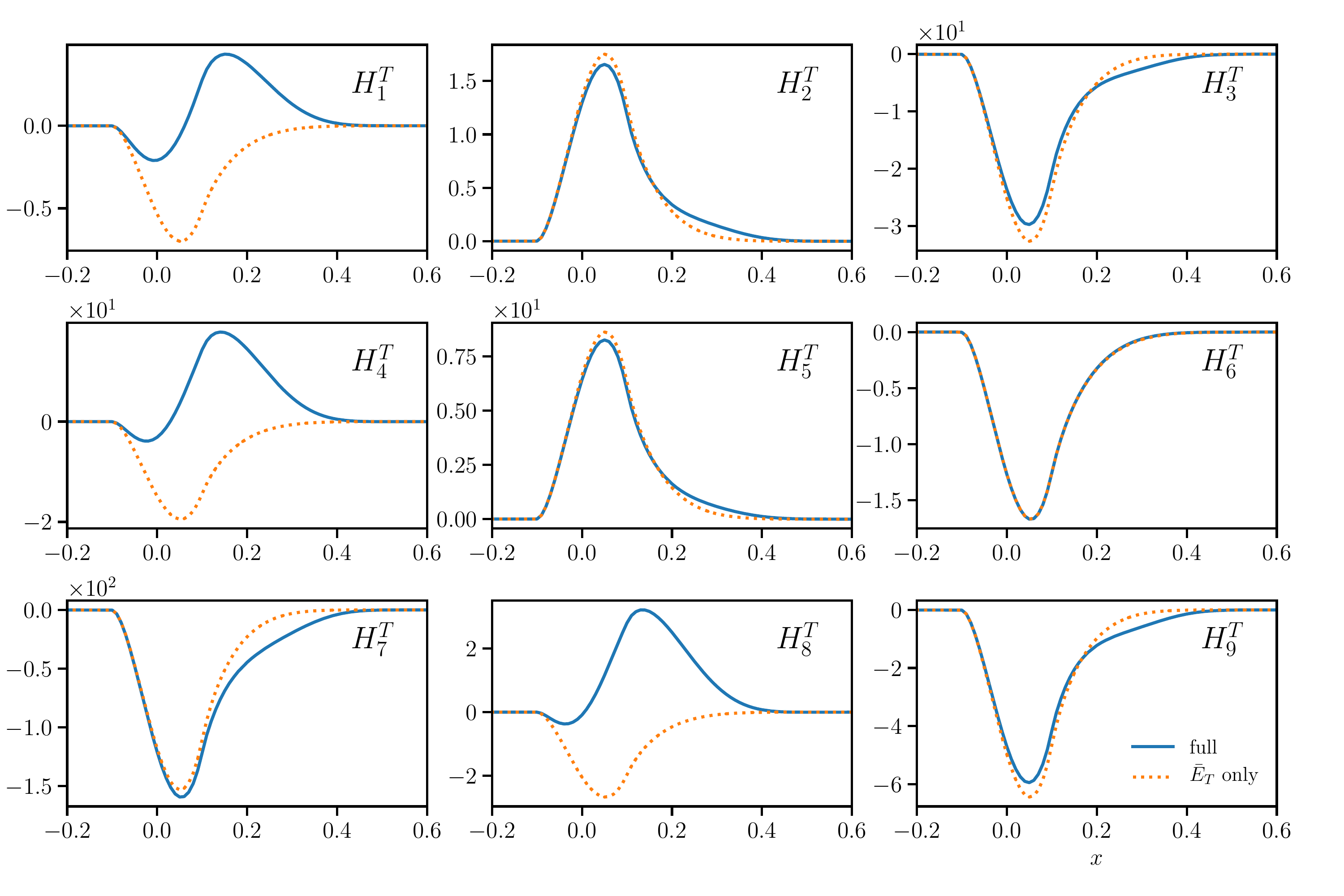}
 \end{center}
 \caption{
 \label{fig:gpds_ht}
(Color online) Deuteron quark transversity GPDs computed in the convolution formalism, at 
$\xi=0.1$,$t=-0.25$~GeV$^2$. 
Dashed curve is a calculation only considering 
the $\bar{E}_T$ nucleon GPD.}
\label{fig:gpd_noht}
\end{figure}

Figures~\ref{fig:amps_ht} and \ref{fig:gpds_ht} show that most helicity 
amplitudes are dominated by the $\bar{E}_T=2\widetilde{H}_T+E_T$ nucleon chiral odd GPD from the GK 
parametrization.  Only the $\mc{A}_{++;0-}$ and $\mc{A}_{0+;--}$ receive large 
contributions from $H_T$.  The dominance of $\bar{E}_T$ in most amplitudes is 
caused by its size on the one hand (which is larger than $H_T$) and the fact 
that both $u$ and $d$ quarks have same sign $\bar{E}_T$ GPDs, whereas they have opposite 
for $H_T$ and thus are small for the isosinglet contribution entering in the 
convolution formula.

\begin{figure}[htb]
\begin{center}
 \includegraphics[width=0.8\textwidth]{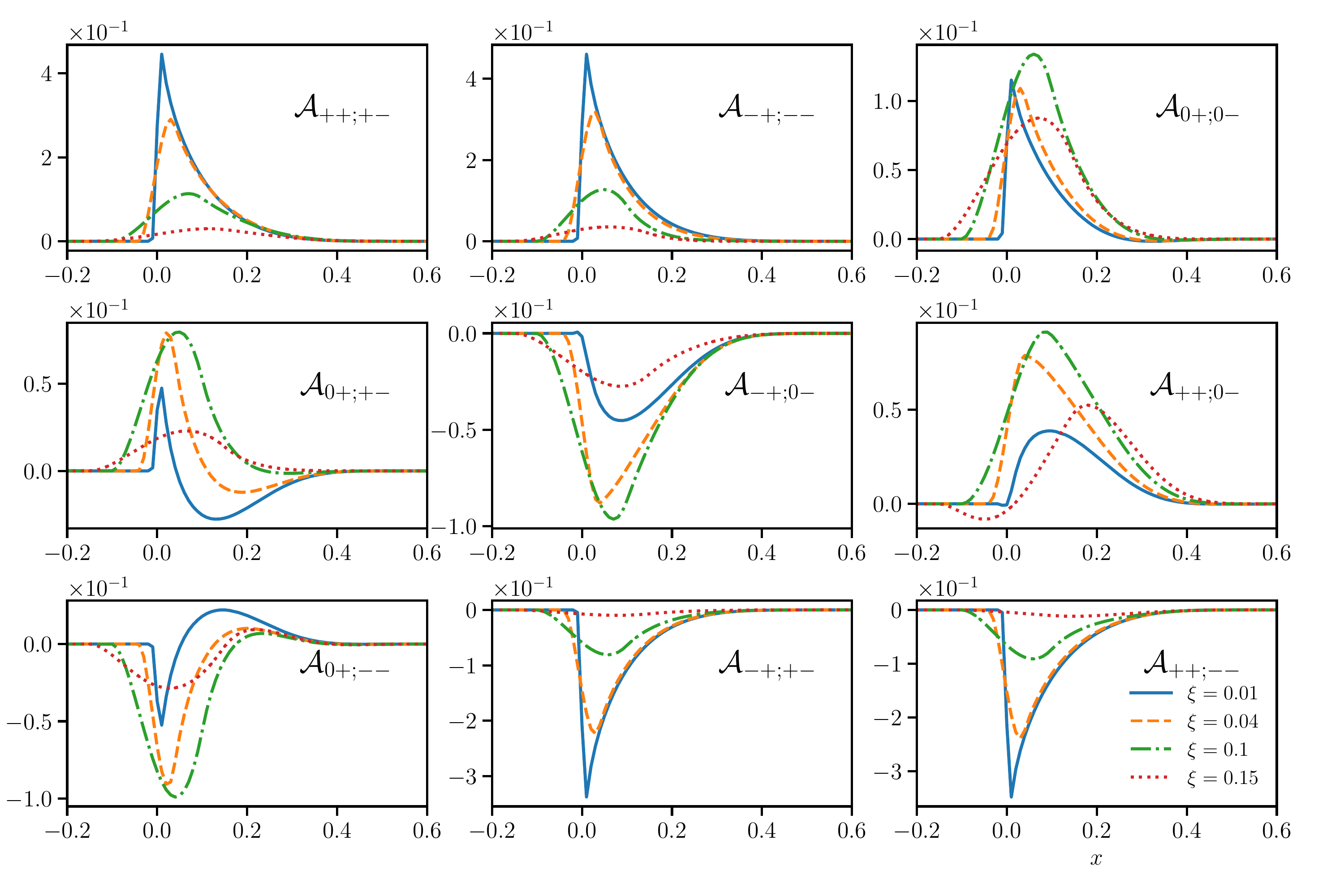}
 \end{center}
 \caption{
 \label{fig:amps_xi}
(Color online) Deuteron quark helicity amplitudes computed in the convolution formalism, at 
various $\xi$ for $t=-0.4$~GeV$^2$.}
\end{figure}

\begin{figure}[htb]
\begin{center}
 \includegraphics[width=0.8\textwidth]{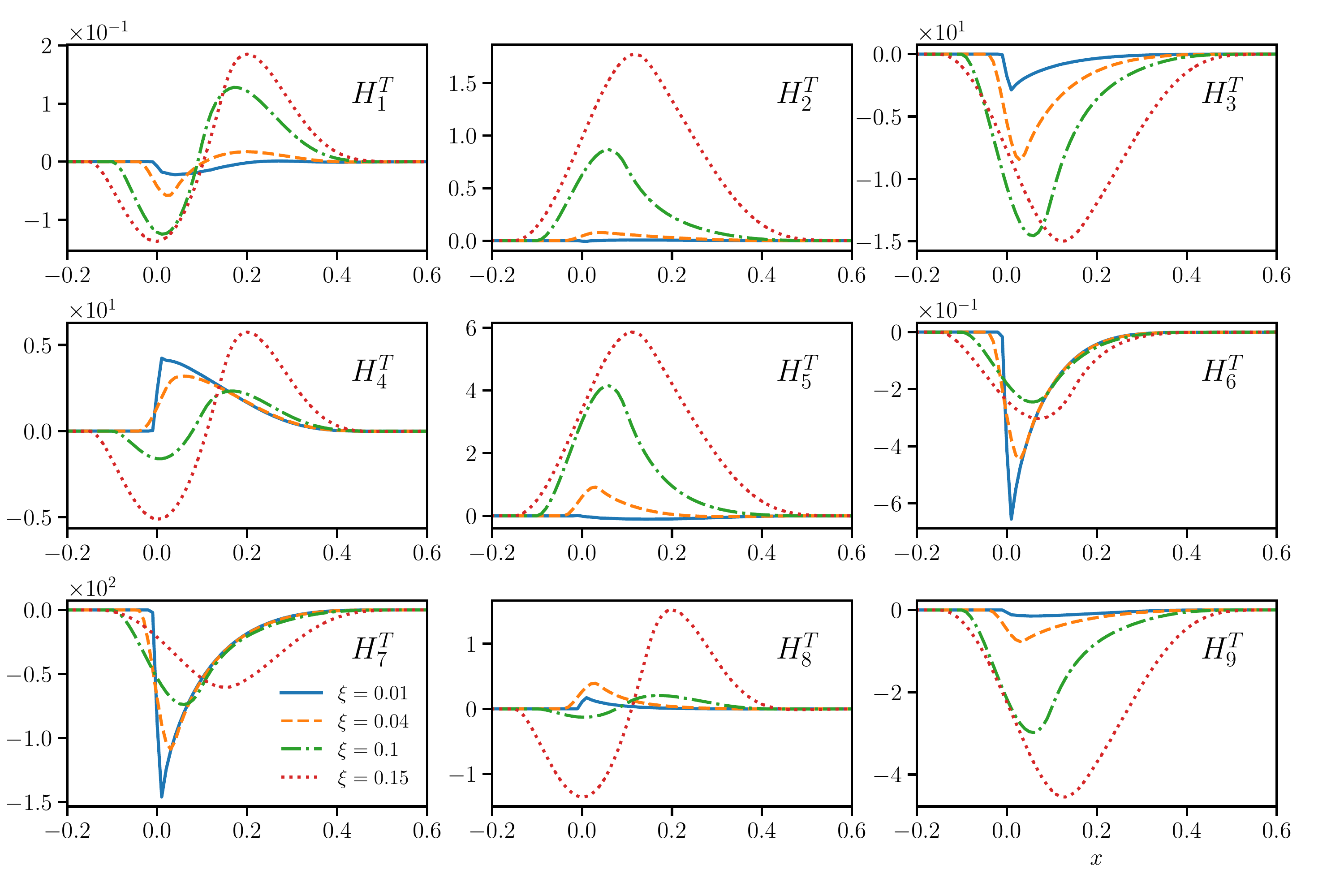}
 \end{center}
 \caption{
 \label{fig:gpds_xi}
(Color online) Deuteron quark transversity GPDs computed in the convolution formalism, at 
various $\xi$ for $t=-0.4$~GeV$^2$.}
\end{figure}

Figs.~\ref{fig:amps_xi} and \ref{fig:gpds_xi} show the $\xi$ dependence of the 
amplitudes and GPDs at a momentum transfer of $t=-0.4~\text{GeV}^2$.  The deuteron helicity amplitudes with zero or two units of deuteron helicity flip decrease significantly with larger $\xi$, while the ones with one unit of helicity flip are largest at intermediate values of $\xi$.  For the GPDs, $H^T_6, H^T_7$ decrease significantly with for larger $\xi$.

%
%
%


\section{Sum rules in the deuteron convolution picture}
\label{sec:sumrules}

\begin{figure}[htb]
\begin{center}
 \includegraphics[width=0.4\textwidth]{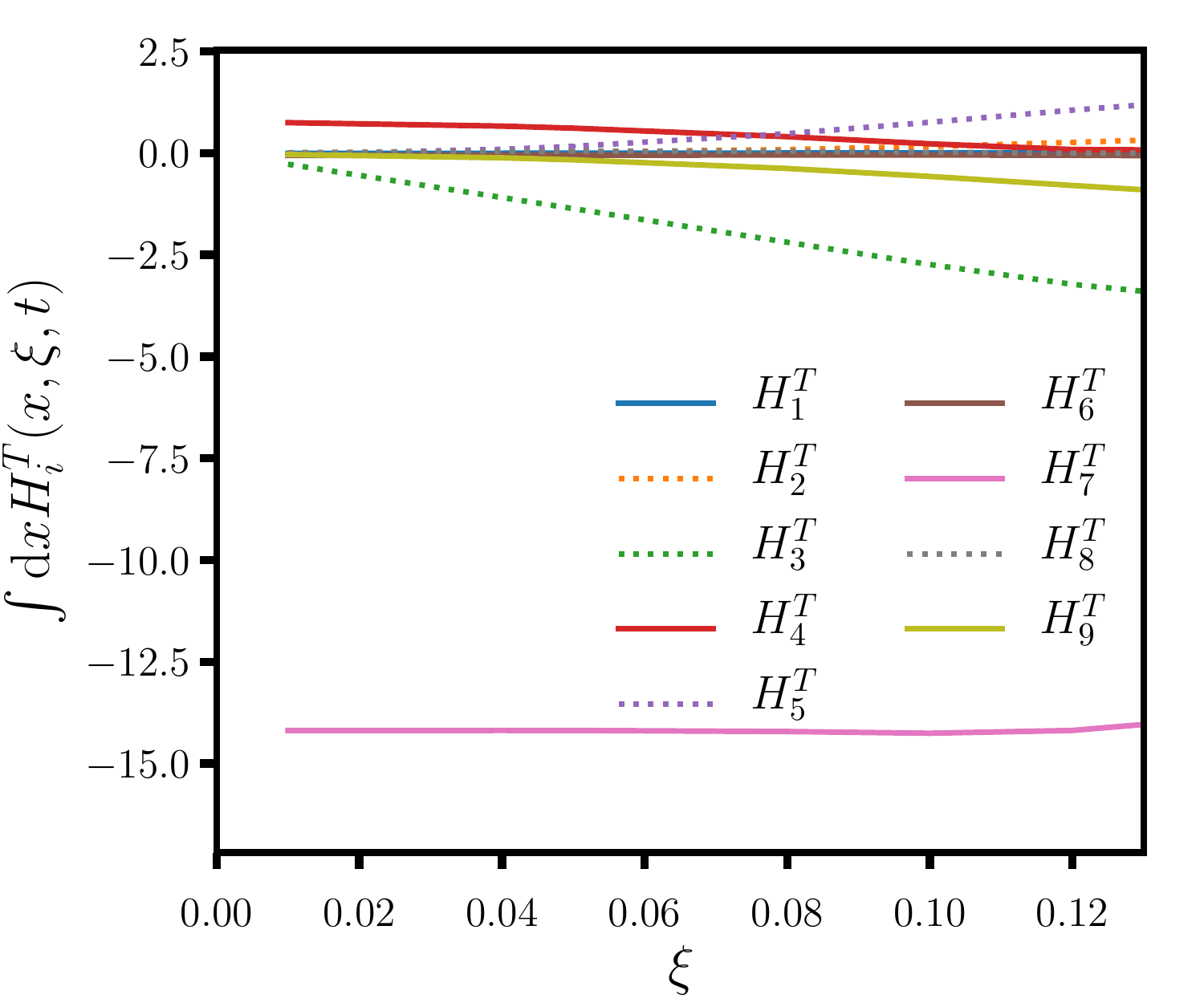}
 \end{center}
 \caption{
 \label{fig:sumrules}
(Color online) First moments of the chiral odd quark GPDs at $t=-0.4~\text{GeV}^2$ as a 
function of skewness $\xi$.  Dashed curves are the GPDs that have zero sum rules.  The moderate $\xi$-dependence  is a consequence of Lorentz symmetry breaking due to lowest order approximation of the convolution picture.
}
\end{figure}

In this section we focus on the quark transversity GPD sum rules of 
Eq.~(\ref{eq:q_sums}).  Because of Lorentz invariance, the GPDs obey polynomiality properties that in particular predict that these first moments should be independent of the value of skewness. As we use a lowest order Fock space expansion in our convolution model, and this explicitly breaks Lorentz invariance (no negative energy projections are included for instance), we  investigate to which degree the $\xi$ independence is  violated in our convolution formalism. Fig.~\ref{fig:sumrules} depicts the 
results for the first moments of all the chiral odd quark GPDs at $t=-0.4~\text{GeV}^2$ (which requires $|\xi|<0.17$).    We see that several GPDs show a significant $\xi$ dependence,
especially the GPDs $H^T_3,H^T_4,H^T_5$ and $H^T_9$. Two of these ($H^T_3,H^T_5$) even should have zero first moments according to 
Eq.~(\ref{eq:q_sums}).  This could be 
seen as a requirement to include higher order contributions in the convolution 
picture, i.e. beyond the handbag diagram or including higher Fock states.  

To investigate this further, we look at the sum rules in a \emph{minimal} 
convolution picture, detailed in App.~\ref{app:min_conv}.  This minimal convolution picture allows us to 
calculate the deuteron GPDs analytically.  Looking at the final expressions for 
the deuteron GPDs listed in Eqs.~\ref{eq:d_gpd_min_conv},
we see that only GPDs $H^T_4,H^T_5,H^T_6$ have a leading term 
$\mc{O}(\xi^0)$.   Inspecting Eqs.~\ref{eq:d_gpd_min_conv}, almost all GPDs have dominating terms proportional to 
$D^{-2}$ (which is large for the deuteron kinematics considered here) that go as higher powers of $\xi$, especially the GPDs 
that also show the largest $\xi$ dependence in the full convolution model.  It 
is worth noting that the fact that $H^T_7=0$ in this minimal convolution is due 
to the lack of a $D$-wave part in the deuteron wave function in this model and not a 
reflection of a sum rule.

The violation of the sum rules thus is an inherent feature of all convolution models based on a Fock space expansion, even the simplest ones. One should thus blame their formulation for Lorentz invariance breaking. The contribution of higher Fock states is beyond the current scope of our study.  One possible approach for the deuteron that respects Lorentz invariance (and thus polynomiality of the GPDs) is the use of the covariant Bethe-Salpeter equation for the deuteron. Current deuteron GPD implementations of the Bethe-Salpeter approach are limited to a contact $NN$ interaction~\cite{Freese:2018priv}, while the approach presented here allows the use of realistic deuteron wave function parametrizations.


\section{Conclusion}
\label{sec:concl}
Our study completes the description of leading twist  quark and and gluon GPDs in the deuteron, in a convolution model based on the impulse approximation 
and using the lowest Fock space state for the deuteron in terms of nucleons. Although this picture is far from complete, it is a necessary starting point for the study 
of exclusive hard reactions in the QCD collinear factorization framework. It will enable us to confront this framework to near future experimental results. We showed that the GPDs were not very model-sensitive to the nucleon nucleon potential as far as the impulse approximation was used. However a richer structure as those involving a hidden color component \cite{Miller:2013hla} should lead to quite different GPDs, in  particular in the gluonic sector.

The transversity sector is remarkably quite difficult to access in hard reactions \cite{Barone:2001sp}, mostly because of the chiral-odd character of the quark transversity distributions. As far as transversity quark GPDs are concerned, the fact that they do not contribute to the leading twist amplitude for the electroproduction of one meson   \cite{Diehl:1998pd, Collins:1999un} lead to the study of higher twist \cite{Ahmad:2008hp,Goloskokov:2011rd,Goldstein:2012az} or quark mass sensitive \cite{Pire:2015iza, Pire:2017lfj} contributions, and to the study of other reactions with more particles in the final state \cite{Ivanov:2002jj, Enberg:2006he, Beiyad:2010cxa, Boussarie:2016qop}. The case for gluon transversity GPDs is rather different since they appear in the leading twist DVCS  \cite{Belitsky:2000jk} and timelike Compton scattering \cite{Berger:2001xd} amplitudes.

We shall address the rich phenomenology of these reactions on the deuteron in future works, both for moderate energy range of JLab~\cite{Armstrong:2017wfw} and for the very high energy range aimed at the EIC \cite{Boer:2011fh} and the LHeC \cite{AbelleiraFernandez:2012cc} with deuteron beams.

\acknowledgments
We acknowledge useful discussions with Adam Freese, C\'edric Lorc\'e, Claude Roiesnel, Lech Szymanowski and Jakub Wagner.  We thank Jakub Wagner for help with the numerical implementation of the chiral odd nucleon GPD parametrization.
\appendix
\section{Conventions} \label{app:conv}
This appendix summarizes the conventions and notations used throughout the text.
We work with the following light-front conventions:
\begin{itemize}
\item Light-front components and Levi-Civita tensor
\begin{align}
 &x^\pm=x^0\pm x^3\,,\nonumber\\
  &\epsilon^{0123}=1\,.
\end{align}

\item We use the transverse $R/L$ indices defined as
\begin{align}\label{eq:RL}
 &a^R=a^x+ia^y\,,\nonumber\\
  &a^L=a^x-ia^y\,,
\end{align}
and for the action of light-front discrete symmetries we need the notation
\begin{equation} \label{eq:tilde}
 \tilde{a}^\mu = (a^+,a^-,-a^1,a^2)\,.
\end{equation}
We have
\begin{align}
 &\tilde{a}^R=-a^L\,,
 \nonumber\\
  &\tilde{a}^L=-a^R\,.
\end{align}
The product of two four-vectors can be written as
\begin{equation}
 a^\mu b_\mu=\tfrac{1}{2}(a^+b^- +a^-b^+ -a^Rb^L-a^Lb^R)\,.
\end{equation}

 \item Single-particle state normalization of light-front helicity states
 \begin{equation}\label{eq:sp_norm}
  \braket{p'\,\lambda'}{p\,\lambda}=(2\pi)^3 
2p^+\delta_{\lambda\lambda'}\delta^{+\perp}(p-p')\,.
 \end{equation}

\item Creation and annihilation operators are normalized through
\begin{equation}
[a^\lambda_p,a^{\lambda'\dagger}_{p'}]_\pm=(2\pi)^3\delta_{\lambda\lambda'}
\delta^{+\perp}(p-p')\,.
\end{equation}

\item The last two equations imply
\begin{equation}
 \ket{p\,\lambda}=\sqrt{2p^+}a^{\dagger\lambda}_p \ket{0}\,.
\end{equation}
\item The Dirac field in light-front quantization becomes
\begin{equation}
 \psi(x)=\sum_{\lambda=\pm \tfrac{1}{2}} \int 
\frac{dk^{+\perp}}{\sqrt{2k^+}(2\pi)^3}\left[a^\lambda_k 
u(k,\lambda)e^{-ikx}+b^{\dagger\lambda}_k v(k,\lambda) e^{ikx} \right]\,,
\end{equation}
with $u(k,\lambda),v(k,\lambda)$ the standard light-front 
spinors~\cite{Brodsky:1997de}.

\item The gluon field (with an implicit summation over a color index and 
$SU(3)$ generators implied)
\begin{equation}
 A^\mu(x)=\sum_{\lambda=\pm}\int 
\frac{dk^{+\perp}}{\sqrt{2k^+}(2\pi)^3}\left[a^\lambda_k 
\epsilon^\mu(k,\lambda) e^{-ikx}+a^{\dagger\lambda}_k 
\epsilon^{\mu*}(k,\lambda) e^{ikx} \right]\,, 
\end{equation}
where the polarization four-vectors are
\begin{align} \label{eq:pol_vectors_g}
&\epsilon^\mu(k,+)=\kbordermatrix{&+&-&1&2\\ &0
&-\frac{\sqrt{2}k^R}{k^+}&-\frac{1}{\sqrt{2 
}}&-\frac{i}{\sqrt{2}} }
\,,\nonumber\\
&\epsilon^\mu(k,-)=\kbordermatrix{&+&-&1&2\\ &0
&\frac{\sqrt{2}k^L}{k^+}&\frac{1}{\sqrt{2 
}}&-\frac{i}{\sqrt{2}} }\,.
\end{align}
Finally, the field strength and dual field strength are
\begin{align}
& G^{\mu\nu}(x)=\partial^\mu A^\nu(x) - \partial^\nu A^\mu(x) 
-ig[A^\mu(x),A^\nu(x)]
&\widetilde{G}^{\mu\nu}(x)=-\frac{1}{2}\epsilon^{\mu\nu\rho\sigma}G_{\rho\sigma}
\,.
\end{align}
\end{itemize}


\section{Light-front discrete symmetries}
\label{app:discrete}
Light-front discrete symmetries were first considered in Ref.~\cite{Soper:1972xc} and are discussed in several other instances of the literature with slightly 
different forms of the operators between them (see for instance in Refs.~\cite{Carlson:2003je,Brodsky:2006ez,Lorce:2013pza}).  We follow the definitions used in 
Refs.~\cite{Brodsky:2006ez,Lorce:2013pza} as the combination of light-front parity and time reversal with the 
standard charge conjugation is consistent with the instant form $CPT$.  To our knowledge, the action of these light-front discrete symmetry operators on single-particle states and quark and gluon fields has not been summarized in detail or in the case it has been written out~\cite{Brodsky:2006ez}, the intermediary formulas contain a number of errors and inconsistencies.  We therefore include a summary here as a pedagogical appendix.

\subsection{Light-front parity}

We can introduce the light-front parity symmetry transformation by its action on a coordinate four-vector:
\begin{equation}
 \Lambda(\mc{P}_\perp): x^\mu \mapsto \tilde{x}^\mu = (x^+,x^-,-x^1,x^2)\,.
\end{equation}
As an operator there are a few possible choices to implement this transformation.   These differ 
in an overall sign of the phase in the exponential, but do not yield differences when considering the action of $\mc{P}_\perp$ on correlator matrix elements.  We choose 
\begin{equation}\label{eq:P_lf_def}
 \mc{P}_\perp = e^{-i\pi J_1}\mc{P} = e^{-i\frac{\pi}{2} J_3}e^{i\pi 
J_2}e^{i\frac{\pi}{2} J_3}\mc{P}\,,
\end{equation}
with $\mc{P}$ the standard instant form parity operator.  We first consider the 
massive case.  In the rest frame, acting with $\mc{P}_\perp$ on a massive single-particle state~\footnote{The normalization of particle states and fields is given in App.~\ref{app:conv}.} with spin $j$ yields using Eq.~(\ref{eq:P_lf_def})
\begin{equation}
 \mc{P}_\perp \ket{j\,m} =  e^{-i\pi j}\eta \ket{j\,m}\,,
\end{equation}
where $\eta$ is the intrinsic parity of the particle.  For light-front helicity states (defined 
with the standard light-front boosts) and using the commutation relations of the Lorentz group algebra, we obtain
\begin{equation}
 \mc{P}_\perp \ket{p\,\lambda}=e^{-i\pi j}\eta \ket{\tilde{p}\,{-\lambda}}\,.
\end{equation}
Light-front parity thus flips the light-front helicity of the particle and transforms its momentum.  For the creation and 
annihiliation operators we obtain
\begin{align}\label{eq:P_lf_a}
 &\mc{P}_\perp a^{\dagger\lambda}_p \mc{P}_\perp^\dagger = \eta_a e^{-i\pi j} 
 a^{\dagger -\lambda}_{\tilde{p}} \,,\nonumber\\
 &\mc{P}_\perp a^{\lambda}_p 
\mc{P}_\perp^\dagger = \eta^*_a e^{i\pi j} 
 a^{-\lambda}_{\tilde{p}} 
\end{align}
For the Dirac field, we have
\begin{equation}
\mc{P}_\perp \psi(x) \mc{P}_\perp^\dagger = \sum_{\lambda=\pm\tfrac{1}{2}} \int
\frac{d\tilde{k}^{+\perp}}{\sqrt{2\tilde{k}^+}(2\pi)^3}
\left[\eta_a^* e^{i\pi j}a^{-\lambda }_{\tilde{k}} 
u(k,\lambda) e^{-i\tilde{k}\tilde{x}}+\eta_b e^{-i\pi j} 
b^{\dagger -\lambda}_{\tilde{k}} v(k,\lambda) e^{i\tilde{k}\tilde{x}} 
\right]\,.
\end{equation}
The light-front spinors have
\begin{align}
 &\gamma^1\gamma_5 u(\tilde{k},-\lambda) = u(k,\lambda)\,, \nonumber\\
 &\gamma^1\gamma_5 v(\tilde{k},-\lambda) = v(k,\lambda)\,,
\end{align}
and when requiring $\eta_b = -\eta_a^*$ as in the instant form case, we have
\begin{align}
& \mc{P}_\perp \psi(x) \mc{P}_\perp^\dagger = \eta_a^* e^{i\pi 
j}\gamma^1\gamma^5\psi(\tilde{x})\,,\nonumber\\
& \mc{P}_\perp \psi^\dagger(x) \mc{P}_\perp^\dagger = \eta_a e^{-i\pi 
j}\psi^\dagger(\tilde{x})\gamma^1\gamma^5\,.
\end{align}
When considering the transformation under $\mc{P}_\perp$ of the quark-quark correlators introduced in Eq.~(\ref{eq:corr_def}), the phases $e^{\pm i\pi j}$ and intrinsic parities do not 
contribute as they cancel between the initial and final state and the two quark 
fields.

For massless states, the little group is characterized by a 
reference state with momentum along the $z$-axis $\bar{k}$ and two possible 
$J_3$ eigenvalues (if parity is a good symmetry).  For this reference state one has
\begin{equation}
 \mc{P}_\perp \ket{\bar{k}\, \lambda}=\eta \ket{\bar{k}\, {-\lambda}}\,.
\end{equation}
For a massless light-front helicity state with general momentum, this yields
\begin{equation}
 \mc{P}_\perp \ket{p\, \lambda}=\eta \ket{\tilde{p}\, {-\lambda}}\,,
\end{equation}
so as in the massive case momentum transforms and light-front helicity flips.  Creation and annihilation 
operators transform as in Eq.~(\ref{eq:P_lf_a}) but without the phase factor.  For the gluon field this 
yields
\begin{equation}
\mc{P}_\perp A^\mu(x) \mc{P}_\perp^\dagger = \sum_{\lambda=\pm} \int
\frac{d\tilde{k}^{+\perp}}{\sqrt{2\tilde{k}^+}(2\pi)^3}
\left[\eta_a^* a^{-\lambda }_{\tilde{k}} 
\epsilon^\mu(k,\lambda) e^{-i\tilde{k}\tilde{x}}+\eta_a 
a^{\dagger -\lambda}_{\tilde{k}} \epsilon^{\mu*}(k,\lambda) 
e^{i\tilde{k}\tilde{x}} 
\right]\,. 
\end{equation}
The polarization four-vectors of Eq.~(\ref{eq:pol_vectors_g}) have
\begin{equation}
 \epsilon^\mu(k,\pm)=\tilde{\epsilon}^{\mu}(\tilde{k},\mp)\,,
\end{equation}
and with $\eta_a$ real we have for the gluon field and field strength
\begin{align}
& \mc{P}_\perp A^\mu(x) \mc{P}_\perp^\dagger=\eta_a 
\widetilde{A}^{\mu}(\tilde{x})\,,\nonumber\\
& \mc{P}_\perp G^{\mu\nu}(x) \mc{P}_\perp^\dagger = \eta_a 
\bar{G}^{\mu\nu}(\tilde{x})\,,
\end{align}
where $\bar{G}^{\mu\nu}(\tilde{x})=G^{\mu\nu}(\tilde{x})$ for an even 
number of indices $1$, and with a minus sign for an odd number.
As in the quark case, the intrinsic parities and phases $e^{\pm i \pi j}$  cancel  in the light-front parity transformation of the gluon-gluon correlators of Eq.~(\ref{eq:corr_def_g}).

\subsection{Time reversal}
We can introduce the time reversal symmetry transformation by its action on a coordinate four-vector :
\begin{equation}
 \Lambda(\mc{T}_\perp): x^\mu \mapsto -\tilde{x}^\mu = (-x^+,-x^-,x^1,-x^2)\,.
\end{equation}
Because of the anti-unitarity of $\mc{T}_\perp$ momenta transform as
\begin{equation}
  p^\mu \mapsto \tilde{p}^\mu = (p^+,p^-,-p^1,p^2)\,.
\end{equation}
As with $\mc{P}_\perp$ there are several choices to write $\mc{T}_\perp$ on the operator level, with no difference at the level of transformation of correlator matrix elements.  We take
\begin{equation} \label{eq:T_lf_def}
 \mc{T}_\perp = e^{-i\pi J_1}\mc{T} = e^{-i\frac{\pi}{2} J_3}e^{i\pi 
J_2}e^{i\frac{\pi}{2} J_3}\mc{T}\,,
\end{equation}
where $\mc{T}$ is the standard instant form time reversal operator.  For 
massive
particles, we have  in the rest frame for a spin $j$ particle using Eq.~
(\ref{eq:T_lf_def})
\begin{equation}
 \mc{T}_\perp \ket{j\,m} =  e^{-i\pi m}\tilde{\eta} \ket{j\,m}\,,
\end{equation}
with $\tilde{\eta}$ a phase.  For light-front helicity states 
we obtain
\begin{equation}
 \mc{T}_\perp \ket{p\,\lambda}=e^{-i\pi \lambda}\tilde{\eta} 
\ket{\tilde{p}\, \lambda}\label{eq:trev}\,.
\end{equation}
Consequently light-front time reveral does not flip light-front helicity, but momentum transforms.  For the creation and 
annihiliation operators we obtain
\begin{align}\label{eq:T_lf_a}
 &\mc{T}_\perp a^{\dagger\lambda}_p \mc{T}_\perp^\dagger = \tilde{\eta}_a 
e^{-i\pi \lambda} 
 a^{\dagger \lambda}_{\tilde{p}} \,,\nonumber\\
 &\mc{T}_\perp a^{\lambda}_p 
\mc{T}_\perp^\dagger = \tilde{\eta}^*_a e^{i\pi \lambda} 
 a^{\lambda}_{\tilde{p}}\,,
\end{align}
and for the Dirac field one has
\begin{equation}
\mc{T}_\perp \psi(x) \mc{T}_\perp^\dagger = \sum_{\lambda=\pm \tfrac{1}{2}} \int
\frac{d\tilde{k}^{+\perp}}{\sqrt{2\tilde{k}^+}(2\pi)^3}
\left[\tilde{\eta}_a^* e^{i\pi \lambda}a^{\lambda }_{\tilde{k}} 
u^*(k,\lambda) e^{-i\tilde{k}(-\tilde{x})}+\tilde{\eta}_b e^{-i\pi \lambda} 
b^{\dagger \lambda}_{\tilde{k}} v^*(k,\lambda) e^{i\tilde{k}(-\tilde{x})} 
\right]\,.
\end{equation}
The light-front spinors have
\begin{align}
 &-\gamma^1\gamma^2 u(\tilde{k},\lambda) = e^{i\pi\lambda} u^*(k,\lambda)\,, \nonumber\\
 &-\gamma^1\gamma^2 v(\tilde{k},\lambda) = e^{-i\pi\lambda} v^*(k,\lambda)\,,
\end{align}
and when requiring $\tilde{\eta}_b = \tilde{\eta}_a^*$ as in the instant form 
case, we arrive at
\begin{align}
& \mc{T}_\perp \psi(x) \mc{T}_\perp^\dagger = \tilde{\eta}_a^* 
(-\gamma^1\gamma^2)\psi(-\tilde{x})\,,\nonumber\\
& \mc{T}_\perp \psi^\dagger(x) \mc{T}_\perp^\dagger = \tilde{\eta}_a 
\psi^\dagger(-\tilde{x})(-\gamma^2\gamma^1)\,.
\end{align}
In the transformation under $\mc{T}_\perp$ of the quark-quark correlators of Eq.~(\ref{eq:corr_def}), all the phases cancel, but there remains an $(-1)^{\lambda'-\lambda}$ factor originating from from the transformation of the
initial and final state [Eq.~(\ref{eq:trev})].

For the massless case, we have for the reference state
\begin{equation}
 \mc{T}_\perp \ket{\bar{k}\,\lambda} =  \tilde{\eta} \ket{\bar{k}\,\lambda}\,,
\end{equation}
and for the massless light-front helicity states with general momentum $p$
\begin{equation}
 \mc{T}_\perp \ket{p\,\lambda} =  \tilde{\eta} \ket{\tilde{p}\,\lambda}\,,
\end{equation}
Also in the massless case, light-front time reversal conserves light-front helicity and momentum is transformed.  
Creation and annihilation 
operators transform as in Eq.~(\ref{eq:T_lf_a}) but without the phase factor.  
For the transformation of the gluon field, we arrive at
\begin{equation}
\mc{T}_\perp A^\mu(x) \mc{T}_\perp^\dagger = \sum_\lambda \int
\frac{d\tilde{k}^{+\perp}}{\sqrt{2\tilde{k}^+}(2\pi)^3}
\left[\tilde{\eta}_a^* a^{\lambda }_{\tilde{k}} 
\epsilon^{\mu*}(k,\lambda) e^{-i\tilde{k}(-\tilde{x})}+\tilde{\eta}_a 
a^{\dagger \lambda}_{\tilde{k}} \epsilon^{\mu}(k,\lambda) 
e^{i\tilde{k}(-\tilde{x})} 
\right]\,. 
\end{equation}
The polarization four-vectors of Eq.~(\ref{eq:pol_vectors}) have
\begin{equation}
 \epsilon^{\mu*}(k,\pm)=-\tilde{\epsilon}^{\mu}(\tilde{k},\pm)\,,
\end{equation}
and with $\tilde{\eta}_a$ real we have for the gluon field and field strength
\begin{align}
& \mc{T}_\perp A^\mu(x) \mc{T}_\perp^\dagger=-\tilde{\eta}_a 
\widetilde{A}^{\mu}(-\tilde{x})\,,\nonumber\\
& \mc{T}_\perp G^{\mu\nu}(x) \mc{T}_\perp^\dagger = -\tilde{\eta}_a 
\bar{G}^{\mu\nu}(-\tilde{x})\,.
\end{align}
When considering the transformation of the gluon-gluon correlators of Eq.~(\ref{eq:corr_def_g}) with $\mc{T}_\perp$, the phases drop out 
but a factor $(-1)^{\lambda'-\lambda}$ remains from the transformation of the 
initial/final state.

\section{Explicit relations between transversity GPDs and helicity amplitudes}
\label{app:helamps}

In the quark sector, the helicity amplitudes $\mc{A}^q_{\lambda'+;\lambda-}$
can be written as a function of the 9 transversity GPDs using Eq.~(\ref{eq:hel_def}):
\begin{align}\label{eq:hel_first}
\mc{A}^q_{++;+-}= -e^{i\phi}D\left[\frac{\xi}{1-\xi}(H^{qT}_3-H^{qT}_4)+H^{qT}_6
+\frac{D^2}{2(1-\xi^2)}H^{qT}_7
+\frac{1}{2(1-\xi)}(H^{qT}_8-H^{qT}_9)
\right]
\end{align}
\begin{align}
\mc{A}^q_{0+;0-}= e^{i\phi}D\left[ 
-\frac{1}{2\sqrt{2}}(H^{qT}_1+\xi H^{qT}_2) 
+ \frac{2\xi}{1-\xi^2} H^{qT}_3
+\frac{2D^2}{1-\xi^2}H^{qT}_4\right.\nonumber\\
+
\frac{1}{\sqrt{2}}H^{qT}_5
+\frac{2D^2-1-\xi^2}{1-\xi^2}H^{qT}_6
+\frac{D^4-\xi^2}{(1-\xi^2)^2}H^{qT}_7
\left.
-\frac{\xi}{1-\xi^2}H^{qT}_8
-\frac{D^2}{1-\xi^2}H^{qT}_9
\right]
\end{align}
\begin{align}
\mc{A}^q_{0+;+-}= \frac{e^{2i\phi}}{\sqrt{2}}\left[ 
\frac{(1-\xi)}{2\sqrt{2}}(H^{qT}_1 -H^{qT}_2)
-\left(D^2\frac{1}{1-\xi}-2\frac{\xi^2}{1-\xi^2} \right)H^{qT}_3
+\left(D^2\frac{3\xi-1}{1-\xi^2}-\frac{2\xi^2}{1-\xi^2} \right)H^{qT}_4
\right.\nonumber\\
+\frac{\xi}{\sqrt{2}}(1-\xi)H^{qT}_5
-\frac{2D^2}{1+\xi}H^{qT}_6
-\frac{D^2}{(1+\xi)(1-\xi^2)}\left(D^2-\xi
 \right)H^{qT}_7\ \nonumber\\
 +\left(-\frac{D^2}{2(1-\xi)}+\frac{\xi}{1-\xi^2} 
\right)H^{qT}_8 
\left. + \left(\frac{(3-\xi)D^2}{2(1-\xi^2)}-\frac{\xi}{1-\xi^2} 
\right)H^{qT}_9
 \right]
\end{align}
\begin{align}
\mc{A}^q_{++;0-}= 
\frac{D^2}{\sqrt{2}(1-\xi)}\left[
	-H^{qT}_3+H^{qT}_4
+2H^{qT}_6
+\frac{D^2+\xi}{1-\xi^2}H^{qT}_7 
 +\frac{1}{2}(H^{qT}_8-H^{qT}_9)\right]
\end{align}
\begin{align}
\mc{A}^q_{-+;+-}= 
e^{3i\phi}\frac{D}{1-\xi^2}\left[2\xi (H^{qT}_3-\xi H^{qT}_4)+ \frac{D^2}{2}H^{qT}_7 
+(\xi H^{qT}_8 -H^{qT}_9)\right]
\end{align}
\begin{align}
\mc{A}^q_{++;--}= 
e^{-i\phi}\frac{D^3}{2(1-\xi^2)}H^{qT}_7 
\end{align}

The other three helicity amplitudes $\mc{A}^q_{\lambda'+;\lambda-}$ can also be 
obtained by applying Eq.~(\ref{eq:TP}) to the ones obtained above:
\begin{align}
\mc{A}^q_{-+;--}= -e^{i\phi}D\left[\frac{\xi}{1+\xi}(H^{qT}_3+H^{qT}_4)+H^{qT}_6
+\frac{D^2}{2(1-\xi^2)}H^{qT}_7
-\frac{1}{2(1+\xi)}(H^{qT}_8+H^{qT}_9)
\right]
\end{align}
\begin{align}
\mc{A}^q_{-+;0-}= \frac{e^{2i\phi}}{\sqrt{2}}\left[ 
\frac{(1+\xi)}{2\sqrt{2}}(H^{qT}_1 +H^{qT}_2)
+\left(D^2\frac{1}{1+\xi}-2\frac{\xi^2}{1-\xi^2} \right)H^{qT}_3
-\left(D^2\frac{3\xi+1}{1-\xi^2}+\frac{2\xi^2}{1-\xi^2} \right)H^{qT}_4
\right.\nonumber\\
-\frac{\xi}{\sqrt{2}}(1+\xi)H^{qT}_5
-\frac{2D^2}{1-\xi}H^{qT}_6
-\frac{D^2}{(1-\xi)(1-\xi^2)}\left(D^2+\xi
 \right)H^{qT}_7 \nonumber\\
 +\left(\frac{D^2}{2(1+\xi)}+\frac{\xi}{1-\xi^2} 
\right)H^{qT}_8
\left. + \left(\frac{(3+\xi)D^2}{2(1-\xi^2)}+\frac{\xi}{1-\xi^2} 
\right)H^{qT}_9
 \right]
\end{align}
\begin{align}\label{eq:hel_last}
\mc{A}^q_{0+;--}= \frac{D^2}{\sqrt{2}(1+\xi)}\left[
	H^{qT}_3+H^{qT}_4
+2H^{qT}_6+\frac{D^2-\xi}{1-\xi^2}H^{qT}_7 
 -\frac{1}{2}(H^{qT}_8+H^{qT}_9)\right]
\end{align}
 
The determinant of the matrix relating the helicity 
amplitudes and the GPDs in the above equations yields
\begin{equation}
\text{Det}_q=-\frac{1}{2^{9/2}} e^{9i\phi}D^{11}\,
,
\end{equation}
which shows that all tensor structures appearing in Eq.~(\ref{eq:corr_def}) 
are linearly independent away from the forward 
limit.

For the gluon helicity amplitudes we obtain largely similar 
expressions as the tensors that are used in the decomposition are very similar. 
 The main differences are (i) the right-hand side of all equations is 
multiplied with an extra 
$e^{i\phi}2D$ factor compared to the quark helicity 
amplitudes and (ii) there are differences for the 
factors multiplying the $H^{qT}_5$ and $H^{qT}_6$ GPDs as a different tensor structure was used:

\begin{align}\label{eq:hel_first_g}
\mc{A}^g_{++;+-}= 
-e^{2i\phi}D^2\left[\frac{2\xi}{1-\xi}(H^{gT}_3-H^{gT}_4)+\frac{D^2}{1-\xi^2}H^{gT}_7
+\frac{1}{1-\xi}(H^{gT}_8-H^{gT}_9)
\right]
\end{align}
\begin{align}
\mc{A}^g_{0+;0-}= e^{2i\phi}2D^2\left[ 
-\frac{1}{2\sqrt{2}}(H^{gT}_1+\xi H^{gT}_2) 
+ \frac{2\xi}{1-\xi^2} H^{gT}_3
+\frac{2D^2}{1-\xi^2}H^{gT}_4
\right.\nonumber\\
+
\frac{1}{4}(H^{gT}_5+H^{gT}_6)
+\frac{D^4-\xi^2}{(1-\xi)^2}
H^{gT}_7
\left.
-\frac{\xi}{1-\xi^2}H^{gT}_8
-\frac{D^2}{1-\xi^2}H^{gT}_9
\right]
\end{align}
\begin{align}
\mc{A}^g_{0+;+-}= e^{3i\phi}\sqrt{2}D\left[ 
\frac{1-\xi}{2\sqrt{2}}(H^{gT}_1 -H^{gT}_2)
-\left(D^2\frac{1}{1-\xi}-2\frac{\xi^2}{1-\xi^2} \right)H^{gT}_3\right.\nonumber\\
+\left(D^2\frac{3\xi-1}{1-\xi^2}-\frac{2\xi^2}{1-\xi^2} \right)H^{gT}_4
-\frac{1}{2}(H^{gT}_5-\xi H^{gT}_6)\nonumber\\
-\frac{D^2}{(1+\xi)(1-\xi^2)}\left(D^2-\xi
 \right)H^{gT}_7
 +\left(-\frac{D^2}{2(1-\xi)}+\frac{\xi}{1-\xi^2} 
\right)H^{gT}_8 \nonumber\\
\left.+  \left(\frac{(3-\xi)D^2}{2(1-\xi^2)}-\frac{\xi}{1-\xi^2} 
\right)H^{gT}_9
 \right]
\end{align}
\begin{align}
\mc{A}^g_{++;0-}= 
e^{i\phi}\frac{\sqrt{2}D^3}{1-\xi}\left[
-H^{gT}_3+H^{gT}_4
+\frac{D^2+\xi}{1-\xi^2}H^{gT}_7 
 +\frac{1}{2}(H^{gT}_8-H^{gT}_9)\right]
\end{align}
\begin{align}
\mc{A}^g_{-+;+-}= \frac{4\xi D^2}{1-\xi^2}(H^{gT}_3-\xi H^{gT}_4)+e^{4i\phi}(H^{gT}_5-\xi^2 H^{gT}_6)+
e^{4i\phi}\frac{D^2}{1-\xi^2}\left[D^2H^{gT}_7 
+2(\xi H^{gT}_8 -H^{gT}_9)\right]
\end{align}
\begin{align}
\mc{A}^g_{++;--}=\frac{D^4}{1-\xi^2}H^{gT}_7 
\end{align}

The other three helicity amplitudes $\mc{A}^g_{\lambda'+;\lambda-}$ can be 
obtained by using Eq.~(\ref{eq:TP}):
\begin{align}
\mc{A}^g_{-+;--}= 
-e^{2i\phi}D^2\left[2\frac{\xi}{1+\xi}(H^{gT}_3+H^{gT}_4)+\frac{D^2}{1-\xi^2}H^{gT}_7
-\frac{1}{1+\xi}(H^{gT}_8+H^{gT}_9)
\right]
\end{align}
\begin{align}
\mc{A}^g_{-+;0-}= e^{3i\phi}\sqrt{2}D\left[ 
\frac{1+\xi}{2\sqrt{2}}(H^{gT}_1 +H^{gT}_2)
+\left(D^2\frac{1}{1+\xi}-2\frac{\xi^2}{1-\xi^2} \right)H^{gT}_3
\right.\nonumber\\
-\left(D^2\frac{3\xi+1}{1-\xi^2}+\frac{2\xi^2}{1-\xi^2} \right)H^{gT}_4
-\frac{1}{2}(H^{gT}_5+\xi H^{gT}_6)\nonumber\\
-\frac{D^2}{(1-\xi)(1-\xi^2)}\left(D^2+\xi
 \right)H^{gT}_7
 +\left(\frac{D^2}{2(1+\xi)}+\frac{\xi}{1-\xi^2} 
\right)H^{gT}_8 \nonumber\\
\left.+  \left(\frac{(3+\xi)D^2}{2(1-\xi^2)}+\frac{\xi}{1-\xi^2} 
\right)H^{gT}_9
 \right]
\end{align}
\begin{align}
\mc{A}^g_{0+;--}= 
e^{i\phi}\frac{\sqrt{2}D^3}{1+\xi}\left[
H^{gT}_3+H^{gT}_4
+\frac{D^2-\xi}{1-\xi^2}H^{gT}_7 
 -\frac{1}{2}(H^{gT}_8+H^{gT}_9)\right]
\end{align}
 
 The determinant of the above set of equations yields
\begin{equation}
\text{Det}_g=-2
e^{18 i \phi }D^{18}\,,
\end{equation}
which is again non-zero away from the forward limit.

For completeness, we also list the inverse relations for both quarks and gluons 
as these are used to obtain the deuteron GPDs from the helicity amplitudes 
calculated in the convolution formalism.

For the quark GPDs we have
\begin{align}\label{eq:ampstoGPD_first}
 H^{qT}_1&=\left[\frac{2\sqrt{2}e^{-i\phi}\xi}{D(1-\xi^2)}\left(\mc{A}^q_{++;+-} 
-\mc{A}^q_{-+;--}\right)
+2e^{-2i\phi}\left(\frac{1}{1-\xi} \mc{A}^q_{0+;+-} 
+\frac{1}{1+\xi}\mc{A}^q_{-+;0-}\right)
 +\frac{2}{1+\xi}\mc{A}^q_{++;0-}
 \right.\nonumber\\
 &\qquad \left.
 +\frac{2}{1-\xi} \mc{A}^q_{0+;--}
 +\frac{2\sqrt{2}D}{(1-\xi^2)}\left(e^{-3i\phi}
\mc{A}^q_{-+;+-} 
-e^{i\phi}\mc{A}^q_{++;--}\right) \right]\,,\\
H^{qT}_2&=
\left[ \frac{\sqrt{2}e^{-i\phi}}{D(1+\xi)^2}\left( \frac{2D^2}{1-\xi}-\xi\right)\mc{A}^q_{++;+-}
-\frac{\sqrt{2}e^{-i\phi}}{D(1-\xi)^2}\left( \frac{2D^2}{1+\xi}+\xi\right)
\mc{A}^q_{-+;--}
+\frac{2\sqrt{2}e^{-i\phi}\xi}{(1-\xi^2)D} \mc{A}^q_{0+;0-}
\right.\nonumber\\
&\qquad -\frac{2e^{-2i\phi}}{1-\xi^2}\left(\mc{A}^q_{0+;+-} - \mc{A}^q_{-+;0-} \right)
+2\left(\frac{1}{(1+\xi)^2}+\frac{2\xi^2}{D^2(1-\xi^2)(1+\xi)} \right)\mc{A}^q_{++;0-} 
\nonumber\\
&\qquad
-2\left(\frac{1}{(1-\xi)^2}+\frac{2\xi^2}{D^2(1-\xi^2)(1-\xi)} \right)\mc{A}^q_{0+;--} 
+\sqrt{2}\frac{e^{-3i\phi}\xi}{D(1-\xi^2)}\mc{A}^q_{-+;+-} 
\nonumber\\
&\qquad \left.
-\frac{\sqrt{2}e^{i\phi}\xi}{D^3(1-\xi^2)^2}\left(4\xi^2+D^2(3+\xi^2) \right)\mc{A}^q_{++;--} 
\right] \,,\\
H^{qT}_3&=\left[-\frac{e^{-i\phi}}{2D}\left(\frac{1-\xi}{1+\xi}  \mc{A}^q_{++;+-}
-\frac{1+\xi}{1-\xi}\mc{A}^q_{-+;--}\right)
-\frac{1}{\sqrt{2}D^2}\left(
 \frac{1-\xi}{1+\xi}\mc{A}^q_{++;0-}-\frac{1+\xi}{1-\xi}\mc{A}^q_{0+;--}\right) 
+\frac{2e^{i\phi}\xi}{D^3(1-\xi^2)} \mc{A}^q_{++;--}\right]\,,\\
H^{qT}_4&=\left[\frac{e^{-i\phi}}{D}\left(\frac{1}{1+\xi}  \mc{A}^q_{++;+-}
+\frac{1}{1-\xi}\mc{A}^q_{-+;--}\right)
+\frac{1}{\sqrt{2}D^2}\left(
 \frac{1-\xi}{1+\xi}\mc{A}^q_{++;0-}+\frac{1+\xi}{1-\xi}\mc{A}^q_{0+;--}\right) 
 +\frac{e^{-3i\phi}}{2D}\mc{A}^q_{-+;+-}
 \right.\nonumber\\
 &\qquad \left. 
-\frac{e^{i\phi}}{2D^3}\left(D^2-\frac{4\xi^2}{1-\xi^2} \right) 
\mc{A}^q_{++;--}\right]\,,\\
H^{qT}_5&=\frac{1}{\sqrt{2}}\left[-\frac{e^{-i\phi}}{D}\left(\frac{1}{(1+\xi)^2}\left(\frac{1}{2}
+\frac{D^2}{1-\xi} \right)\mc{A}^q_{++;+-} 
+\frac{1}{(1-\xi)^2}\left(\frac{1}{2}
+\frac{D^2}{1+\xi} \right)\mc{A}^q_{-+;--} \right)
+\frac{e^{-i\phi}}{(1-\xi^2)D}\mc{A}^q_{0+;0-}
 \right.\nonumber\\
 &\qquad \left.
 +\frac{e^{-2i\phi}}{\sqrt{2}(1-\xi^2)}\left(\mc{A}^q_{0+;+-}+\mc{A}^q_{-+;0-} 
\right)
-\frac{1}{\sqrt{2}(1+\xi)^2}\left(1-\frac{2\xi}{D^2(1-\xi)}\right)\mc{A}^q_{
++;0-}
 \right.\nonumber\\
 &\qquad \left.
-\frac{1}{\sqrt{2}(1-\xi)^2}\left(1+\frac{2\xi}{D^2(1+\xi)}\right)\mc{A}
^q_{0+;--}-\frac{e^{-3i\phi}}{2D(1-\xi^2)}\mc{A}^q_{-+;+-}
-\frac{e^{i\phi}\left(D^2(3\xi^2+1)+4\xi^2 
\right)}{2D^3(1-\xi^2)^2}\mc{A}^q_{++;--}
\right]
\,,\\
\label{eq:HT6}
H^{qT}_6&=-\frac{1}{2D}\left[e^{-i\phi} \left(\mc{A}^q_{++;+-} 
+\mc{A}^q_{-+;--} \right) + e^{-3i\phi}\mc{A}^q_{-+;+-} 
+e^{i\phi}\mc{A}^q_{++;--}\right]\,,\\
H^{qT}_7&=\frac{2e^{i\phi}(1-\xi^2)}{D^3}\mc{A}^q_{++;--}\,,\\
\label{eq:HT8}
H^{qT}_8&=\frac{1}{D}\left[-e^{-i\phi}\left(\frac{1-\xi}{1+\xi}\mc{A}^q_{++;+-} 
-\frac{1+\xi}{1-\xi}\mc{A}^q_{-+;--} \right)
+\frac{\sqrt{2}\xi}{D}\left(\frac{1-\xi}{1+\xi}\mc{A}^q_{++;0-} 
+\frac{1+\xi}{1-\xi}\mc{A}^q_{-+;0-} \right)
+\frac{4e^{i\phi}\xi^3}{D^2(1-\xi^2)} 
\mc{A}^q_{++;--}  \right]\,,\\
H^{qT}_9&=-\frac{1}{D}\left[2e^{-i\phi}\xi\left(\frac{1}{1+\xi}\mc{A}^q_{++;+-} 
-\frac{1}{1-\xi}\mc{A}^q_{-+;--} \right)
+\frac{\sqrt{2}\xi}{D}\left(\frac{1-\xi}{1+\xi}\mc{A}^q_{++;0-} 
-\frac{1+\xi}{1-\xi}\mc{A}^q_{-+;0-} \right)\right.\nonumber\\
&\qquad \left.
+e^{-3i\phi}\mc{A}^q_{-+;+-}
-e^{i\phi}\left(1+\frac{4\xi^2}{D^2(1-\xi^2)} \right)
\mc{A}^q_{++;--}\right]\,.\label{eq:ampstoGPD_last}
\end{align}

For the gluon GPDs we have
\begin{align}
  H^{gT}_1&=\left[\frac{\sqrt{2}e^{-2i\phi}\xi}{D^2(1-\xi^2)}\left(\mc{A}^g_{++;+-} 
-\mc{A}^g_{-+;--}\right)
+\frac{e^{-3i\phi}}{D}\left(\frac{1}{1-\xi} \mc{A}^g_{0+;+-} 
+\frac{1}{1+\xi}\mc{A}^g_{-+;0-}\right)
 +\frac{e^{-i\phi}}{D}\left(\frac{1}{1+\xi}\mc{A}^g_{++;0-}
 \right.\right.\nonumber\\
 &\qquad \left.\left.
 +\frac{1}{1-\xi} \mc{A}^g_{0+;--}\right)
 +\frac{\sqrt{2}}{(1-\xi^2)}\left(e^{-4i\phi}
\mc{A}^g_{-+;+-} 
-\mc{A}^g_{++;--}\right) \right]\,,\\
H^{gT}_2&=\left[-\frac{\sqrt{2}e^{-2i\phi}}{D^2}\left(\frac{1}{(1+\xi)^2}
\left(\xi-D\right) \mc { A } ^g_ { ++;+- } 
+\frac{1}{(1-\xi)^2}\left(\xi+D \right)\mc{A}^g_{-+;--}\right)
+\frac{2\sqrt{2}e^{-2i\phi }\xi}{D^2(1-\xi^2)}\mc{A}^g_{0+;0-} 
\right. \nonumber\\
&\left.\qquad
-e^{-3i\phi}\left(\frac{1}{1+\xi} \mc{A}^g_{0+;+-} 
-\frac{1}{1-\xi}\mc{A}^g_{-+;0-}\right)
+\frac{e^{-i\phi}}{D^3(1-\xi^2)}\left(\frac{4\xi^2+D(1-\xi)^2}{1+\xi}
 \mc{A}^g_{++;0-}
\right.\right. \nonumber\\
&\left.\left.\qquad
 -\frac{4\xi^2+D(1+\xi)^2}{1-\xi} \mc{A}^g_{0+;--} \right)
-\frac{2\sqrt{2}\xi\left(D(1+\xi^2)+2\xi^2 \right)}{D^4(1-\xi^2)^2}\mc{A}^g_{++;--}
\right]\,,\\
H^{gT}_3&=\left[-\frac{e^{-2i\phi}}{4D^2}\left(\frac{1-\xi}{1+\xi}  \mc{A}^g_{++;+-}
-\frac{1+\xi}{1-\xi}\mc{A}^g_{-+;--}\right)
-\frac{e^{-i\phi}}{2\sqrt{2}D^3}\left(
 \frac{1-\xi}{1+\xi}\mc{A}^g_{++;0-}-\frac{1+\xi}{1-\xi}\mc{A}^g_{0+;--}\right) 
+\frac{\xi}{D^4(1-\xi^2)} \mc{A}^g_{++;--}\right]\,,\\
H^{gT}_4&=\left[\frac{e^{-2i\phi}}{4D^2}\left(\frac{1}{1+\xi}  \mc{A}^g_{++;+-}
+\frac{1}{1-\xi}\mc{A}^g_{-+;--}\right)
+\frac{e^{-i\phi}}{2\sqrt{2}D^3}\left(
 \frac{1-\xi}{1+\xi}\mc{A}^g_{++;0-}+\frac{1+\xi}{1-\xi}\mc{A}^g_{0+;--}\right) 
 +\frac{e^{-3i\phi}}{2D}\mc{A}^g_{-+;+-}
 \right.\nonumber\\
 &\qquad \left. 
-\frac{e^{i\phi}}{2D^3}\left(D^2-\frac{4\xi}{1-\xi^2} \right) 
\mc{A}^g_{++;--}\right]\,,\\
H^{gT}_5&=\left[\frac{e^{-2i\phi}}{D(1-\xi^2)}\left(\frac{D(1-\xi)(1+2\xi)+2\xi^3}{
(
1-\xi^2)(1+\xi)}\mc{A}^g_{++;+-} + \frac{D(1+\xi)(1-2\xi)-2\xi^3}{(
1-\xi^2)(1-\xi)}\mc{A}^g_{-+;--} + 2\xi^2 \mc{A}^g_{0+;0-}\right)
\right.\nonumber\\
&\left.\qquad
+\frac{\sqrt{2}e^{-3i\phi}\xi^2}{D(1-\xi^2)}\left( \mc{A}^g_{0+;+-} + 
\mc{A}^g_{-+;0-}\right)
-\frac{\sqrt{2}e^{-i\phi}}{D^3(1-\xi^2)}\left( 
\frac{D(1-\xi)-2\xi}{1+\xi}\mc{A}^g_{++;0-} 
+\frac{D(1+\xi)+2\xi}{1-\xi}\mc{A}^g_{0+;--} \right)
\right.\nonumber\\
&\left.\qquad
+\frac{e^{-4i\phi}}{1-\xi^2}\mc{A}^g_{-+;+-}
+\frac{(D^2(1-\xi^2)-\xi^2)(D^2+4\xi^2)}{D^4(1-\xi^2)^2}\mc{A}^g_{++;--}
\right]\,,\\
H^{gT}_6&=\left[-\frac{e^{-2i\phi}}{D^2(1-\xi^2)}\left(\frac{D(1-\xi)-2\xi}{(
1-\xi^2)(1+\xi)}\mc{A}^g_{++;+-} + \frac{D(1+\xi)+2\xi}{(
1-\xi^2)(1-\xi)}\mc{A}^g_{-+;--} +2\mc{A}^g_{0+;0-} \right) 
\right.\nonumber\\
&\left.\qquad
+\frac{\sqrt{2}e^{-3i\phi}}{D(1-\xi^2)}\left( \mc{A}^g_{0+;+-} + 
\mc{A}^g_{-+;0-}\right)
-\frac{\sqrt{2}e^{-i\phi}\xi^2}{D^3(1-\xi^2)}\left( 
\frac{D(1-\xi)-2\xi}{1+\xi}\mc{A}^g_{++;0-} 
+\frac{D(1+\xi)+2\xi}{1-\xi}\mc{A}^g_{0+;--} \right)
\right.\nonumber\\
&\left.\qquad
+\frac{e^{-4i\phi}}{1-\xi^2}\mc{A}^g_{-+;+-}
+\frac{(D^2(1-\xi)-2\xi)(D^2(1+\xi)+2\xi)}{D^4(1-\xi^2)^2}\mc{A}^g_{++;--}
\right]\,,\\
H^{gT}_7&=\frac{1-\xi^2}{D^4} \mc{A}^g_{++;--}\,,\\
H^{gT}_8&=\frac{1}{D^2}\left[\frac{e^{-2i\phi}}{2}\left(-\frac{1-\xi}{1+\xi}\mc{A}^g_{++;+-}+ 
\frac{1+\xi}{1-\xi}\mc{A}^g_{-+;--} \right)\right.\nonumber\\
&\left.\qquad
+\frac{e^{-i\phi}}{\sqrt{2}D}\left(\frac{1-\xi}{1+\xi}\mc{A}^g_{++;0-}+ 
\frac{1+\xi}{1-\xi}\mc{A}^g_{0+;--} \right)
 + \frac{2\xi^3}{D^2(1-\xi^2)} \mc{A}^g_{++;--}\right]\,,\\
H^{gT}_9&=\frac{1}{D^2}\left[\frac{e^{-2i\phi}}{2}\left(\frac{1-\xi}{1+\xi}\mc{A}^g_{++;+-}+ 
\frac{1+\xi}{1-\xi}\mc{A}^g_{-+;--} \right)\right.\nonumber\\
&\left.\qquad
+\frac{e^{-i\phi}}{\sqrt{2}D}\left(-\frac{1-\xi}{1+\xi}\mc{A}^g_{++;0-}+ 
\frac{1+\xi}{1-\xi}\mc{A}^g_{0+;--} \right)
 - \left(1 + \frac{2\xi^2}{D^2(1-\xi^2)}\right)\mc{A}^g_{++;--}\right]\,.
\end{align}

\section{Minimal convolution model for the deuteron}
\label{app:min_conv}

In this appendix, we outline a minimal convolution model for the deuteron GPDs. 
The model allows to calculate the transversity GPDs analytically and to 
check certain trends seen in the full convolution model.  

The minimal model 
starts from the following assumptions:
\begin{itemize}
 \item We only include the nucleon chiral odd GPD $\bar{E}_T$ and put all others 
equal to zero.  Figs.~\ref{fig:amp_noht} and \ref{fig:gpd_noht} show that this 
is a reasonable starting point.

\item We do not include a $D$-wave component in the deuteron wave function.
\item We do not consider a spatial wave function for the $S$-wave.  This means 
we only include the nucleon spin sums (through Clebsch-Gordan coefficients) and 
consider the following symmetric kinematics in the convolution:  
\begin{align}
&\bm P_\perp=0\,, &\Delta^y=0\,,\nonumber \\ 
&\phi=0\,,\nonumber \\
 &\alpha_1 = 1+\xi\,,  &\alpha'_1 = 1-\xi\,,\nonumber \\ 
 &k_\perp^x=-\frac{\Delta^x}{4}\,, &k_\perp^y=0 \,,
\nonumber \\
 &{k'}_\perp^x=\frac{\Delta^x}{4}\,, &{k'}_\perp^y=0\,,
\nonumber \\
&\xi_N=\frac{2\xi}{1+\xi^2}\,, &x_N=\frac{2x}{1+\xi^2} &\,.
\end{align}
\end{itemize}
With the choice of this kinematics the symmetry constraints of Subsec.~\ref{subsec:gpd_rel} are still obeyed.

In this minimal convolution model, we obtain for the nucleon helicity amplitudes
\begin{align}
 &\int \mathrm{d}x\,_N \mc{A}^N_{++;+-}(x_N,\xi_N,t)= 
(1-\xi_N)\frac{\sqrt{t_{0N}-t}}{4m}F(t)\,,\nonumber\\
&\int \mathrm{d}x\,_N \mc{A}^N_{-+;--}(x_N,\xi_N,t)= 
(1+\xi_N)\frac{\sqrt{t_{0N}-t}}{4m}F(t)\,,\nonumber\\
 &\int \mathrm{d}x\,_N \mc{A}^N_{++;--}(x_N,\xi_N,t)= -
\frac{\xi_N^2}{\sqrt{1-\xi_N^2}}F(t)\,,\nonumber\\
 &\int \mathrm{d}x\,_N \mc{A}^N_{-+;+-}(x_N,\xi_N,t)= 0\,,
\end{align}
where $F(t)=\int \mathrm{d}x\,_N \bar{E}_T(x_N,\xi_N,t)$.

Using Eq.~(\ref{eq:conv_final}) in the minimal version, we obtain for the 
deuteron helicity amplitudes
\begin{align}
 &\int \mathrm{d}x\, \mc{A}^q_{++;+-}(x,\xi,t)= 
(1-\xi)^2\frac{\sqrt{t_{0N}-t}}{2m}F(t)\,,\nonumber\\
 &\int \mathrm{d}x\, \mc{A}^q_{-+;--}(x,\xi,t)= 
(1+\xi)^2\frac{\sqrt{t_{0N}-t}}{2m}F(t)\,,\nonumber\\
&\int \mathrm{d}x\, \mc{A}^q_{0+;0-}(x,\xi,t)= 
(1+\xi^2)\frac{\sqrt{t_{0N}-t}}{2m}F(t)\,,\nonumber\\
&\int \mathrm{d}x\, \mc{A}^q_{0+;+-}(x,\xi,t)= \int \mathrm{d}x\, 
\mc{A}^q_{-+;0-}(x,\xi,t)=0\,,\nonumber\\
&\int \mathrm{d}x\, \mc{A}^q_{++;0-}(x,\xi,t)= \int \mathrm{d}x\, 
\mc{A}^q_{0+;--}(x,\xi,t)=-4\sqrt{2}\frac{\xi^2}{1-\xi^2}F(t)\,,\nonumber\\
&\int \mathrm{d}x\, \mc{A}^q_{++;--}(x,\xi,t)= \int \mathrm{d}x\, 
\mc{A}^q_{-+;+-}(x,\xi,t)=0\,.\nonumber\\
\end{align}
Note that the first moments of $\mc{A}^{0++-}(x,\xi,t)$ and  $\mc{A}^{-+0-}(x,\xi,t)$ 
are zero because the first moment of $\mc{A}_N^{-++-}(x_N,\xi_N,t)$ is 
zero (which 
is the only one contributing to those on the nucleon level), and the first 
moments of $\mc{A}^{++--}(x,\xi,t)$ and $\mc{A}^{-++-}(x,\xi,t)$ are zero because we 
did not include a $D$-wave in the deuteron wave function.

Finally, using Eqs.~(\ref{eq:ampstoGPD_first}) to (\ref{eq:ampstoGPD_last}), we 
obtain for the chiral odd quark GPDs
\begin{align} \label{eq:d_gpd_min_conv}
&\int \mathrm{d}x\,  
H^T_1(x,\xi,t)=-16\sqrt{2}\frac{\xi^2}{(1-\xi^2)^2}\left(\sqrt{\frac{
(1-\xi^2)(t_{0N}-t)
} {(t_0-t) }}\frac{M}{2m}+1 \right)F(t)\,,\nonumber\\
&\int \mathrm{d}x\,  
H^T_2(x,\xi,t)=-4\sqrt{2}\frac{\xi}{(1-\xi^2)^2}\left(1+\frac{\xi^2}{D^2}\right)\left(D(3+\xi^2)\frac{\sqrt{t_{0N}-t}}{2m}+\frac{8\xi^2}{1-\xi^2} \right)
F(t)\,,\nonumber\\
&\int \mathrm{d}x\,  H^T_3(x,\xi,t)=4\xi\frac{(1+\xi^2)}{(1-\xi^2)}
\sqrt{\frac{(t_{0N}-t)}{(t_0-t)(1-\xi^2)}}\frac{M}{m}F(t)
-16\frac{\xi^3}{D^2(1-\xi^2)^2}F(t)\,,\nonumber\\
&\int \mathrm{d}x\,  H^T_4(x,\xi,t)= 
2\frac{(1+3\xi^2)}{(1-\xi^2)}\sqrt{\frac{(t_{0N}-t)}{(t_0-t)(1-\xi^2)}}\frac{M}{
m } F(t) -8\xi^2\frac{1+\xi^2}{D^2(1-\xi^2)^2}F(t)\,,\nonumber\\
&\int \mathrm{d}x\,  
H^T_5(x,\xi,t)=16\sqrt{2}\frac{\xi^4}{D^2(1-\xi^2)^3}F(t)+8\sqrt{2}\xi^2\frac{
(1+\xi^2)}
{(1-\xi^2)^3 
}F(t)-\frac{\sqrt{(1-\xi^2)(t_0-t)(t_{0N}-t)}}{\sqrt{2}Mm}\frac{(1+3\xi^2)}{(1-\xi^2)}
F(t)\nonumber\\
&\qquad\qquad\qquad\qquad
-4\sqrt{\frac{2(t_{0N}-t)}{
(t_0-t)(1-\xi^2) } } 
\frac{M}{2m}\xi^2\frac{(3+\xi^2)}{(1-\xi^2)^2}F(t)\,,\nonumber\\
&\int \mathrm{d}x\,  
H^T_6(x,\xi,t)=-(1+\xi^2)\sqrt{\frac{(t_{0N}-t)}{(t_0-t)(1-\xi^2)}}\frac{M}{m}
F(t)\,, 
\nonumber\\
&\int \mathrm{d}x\,  H^T_7(x,\xi,t)=0\,,\nonumber\\
&\int \mathrm{d}x\,  
H^T_8(x,\xi,t)=16\xi\frac{1+\xi^2}{1-\xi^2}\frac{M}{2m}\sqrt{\frac{t_{0N}-t}{(t_0-t)(1-\xi^2)}}F(t)+16\frac{\xi^3(1+\xi^2)}{D^2(1-\xi^2)^2}F(t)\,,
\nonumber \\
&\int \mathrm{d}x\,  
H^T_9(x,\xi,t)=8\xi^2\frac{3+\xi^2}{1-\xi^2}\frac{M}{2m}\sqrt{\frac{t_{0N}-t}{(t_0-t)(1-\xi^2)}}F(t)-\frac{32\xi^4}{D^2(1-\xi^2)^2}F(t)\,
.
\end{align}


\bibliography{gpd.bib}

\end{document}